\documentclass[preprint,journal]{vgtc}            % preprint (journal style)

\usepackage{xcolor}
\usepackage{listings}

\definecolor{promptgray}{gray}{0.95} % Light gray background

\lstdefinestyle{promptstyle}{
    backgroundcolor=\color{promptgray},
    basicstyle=\ttfamily\small,
    breaklines=true,
    extendedchars=true, % Allows non-ASCII characters 
    literate={-}{{-}}1,  % Ensures hyphens are handled correctly
    frame=single,
    rulecolor=\color{lightgray},
    xleftmargin=10pt,
    xrightmargin=10pt,
    framesep=5pt
}

\usepackage{environ}            % REQUIRED for \NewEnviron
\IfFileExists{ulem.sty}{\usepackage[normalem]{ulem}}{}
\providecommand{\sout}[1]{#1}
\newcommand{\dstrike}[1]{\sout{#1}}

\newif\ifshowchanges
\showchangesfalse

\newif\ifshownotes
\shownotesfalse

\definecolor{addblue}{rgb}{0.00,0.20,0.75}
\definecolor{delred}{rgb}{0.70,0.05,0.05}
\definecolor{notepurple}{rgb}{0.45,0.10,0.55}

\ifshowchanges
  \newcommand{\added}[1]{{\color{addblue}#1}}
  \newcommand{\deleted}[1]{{\color{delred}\dstrike{#1}}}
  \newcommand{\replaced}[2]{{\color{addblue}#1}\,{\color{delred}\dstrike{#2}}}
  \NewEnviron{addedblock}{\begingroup\color{addblue}\BODY\endgroup}
  \NewEnviron{deletedblock}{\begingroup\color{delred}\BODY\endgroup}
\else
  \newcommand{\added}[1]{#1}
  \newcommand{\deleted}[1]{}
  \newcommand{\replaced}[2]{#1}
  \NewEnviron{addedblock}{\BODY}
  \NewEnviron{deletedblock}{}
\fi

\ifshowchanges
  \newcommand{\addedhere}[1]{{\color{addblue}\textsf{\footnotesize[new in this revision: #1]}}\ }
\else
  \newcommand{\addedhere}[1]{}
\fi

\ifshownotes
  \ifshowchanges
    \newcommand{\revnote}[1]{{\color{notepurple}\,$^{\textsf{\scriptsize[#1]}}$}}
  \else
    \newcommand{\revnote}[1]{}
  \fi
\else
  \newcommand{\revnote}[1]{}
\fi

\onlineid{9876}

\preprinttext{}

\vgtccategory{Research}

\title{EVis+: Literature-Grounded Visual Analytics for Hypothesis-Driven Geoscientific Data Exploration}

\author{%
  \parbox{\textwidth}{\centering
    Zhaocong Yang,
    Jing Yang,
    Martha-Cary Eppes,
    Ye Zhao,
    Matan Ben-Asher,
    Sachin Kumar,
    Tinghao Feng,\\
    Deepshikha Bhati,
    Tsung Heng Wu,
    Abe Campbell,
    Lucky Kodwani,
    Ujwal Suresh Vanjare, and
    Surya Kartheek Birudukota
  }
}

\authorfooter{%
    \item
    Zhaocong Yang, Jing Yang, Martha-Cary Eppes,
    Sachin Kumar, Abe Campbell, Lucky Kodwani,
    Ujwal Suresh Vanjare, and Surya Kartheek Birudukota
    are with the University of North Carolina at Charlotte.

    \item
    Ye Zhao, Deepshikha Bhati, and Tsung Heng Wu
    are with Kent State University.

    \item
    Matan Ben-Asher is with The Open University of Israel.

    \item
    Tinghao Feng is with Appalachian State University.

    \item
    Correspondence: \{zyang19, jing.yang, meppes\}@charlotte.edu
}

\abstract{ Scientific researchers often interpret observational data in
light of prior findings reported in the scientific literature. However,
scientific literature typically remains external to visual analytics
environments, requiring researchers to manually connect published findings with
the data and hypotheses under investigation. \replaced{In this paper, we present EVis+, a literature-grounded
visual analytics system for condition–event time series analysis in geoscience.}{To bridge this gap, we propose a visual analytics workflow that leverages large language models (LLMs) to tightly integrate two forms of knowledge.}\revnote{R2b}
EVis+ enhances observational data exploration by (1) retrieving and synthesizing
literature-derived insights relevant to the dataset and analytical goals, (2)
transforming user-selected insights into testable hypotheses and mapping them to
domain-specific visualizations through expert-curated templates, and (3)
supporting evidence examination, discrepancy validation, and literature-driven
re-exploration to support iterative
hypothesis refinement and scientific
exploration. \added{Rather than handing scientific analysis to an
unconstrained language model, EVis+ uses structured AI-assisted transformations
that keep intermediate results visible, allowing researchers to inspect and
guide the analysis at each step.}\revnote{R1, R3.2} \replaced{We evaluate the system through an
evidence-grounding audit, a structured expert assessment of the insights, and
case studies with geoscientists and a comparative study with domain experts.}{Case studies on this prototype with geoscientists analyzing observational datasets illustrate the effectiveness and utility of our workflow.}\revnote{SR: evaluation}

 }


\teaser{
  \centering
  \includegraphics[width=0.80\linewidth, alt={}]{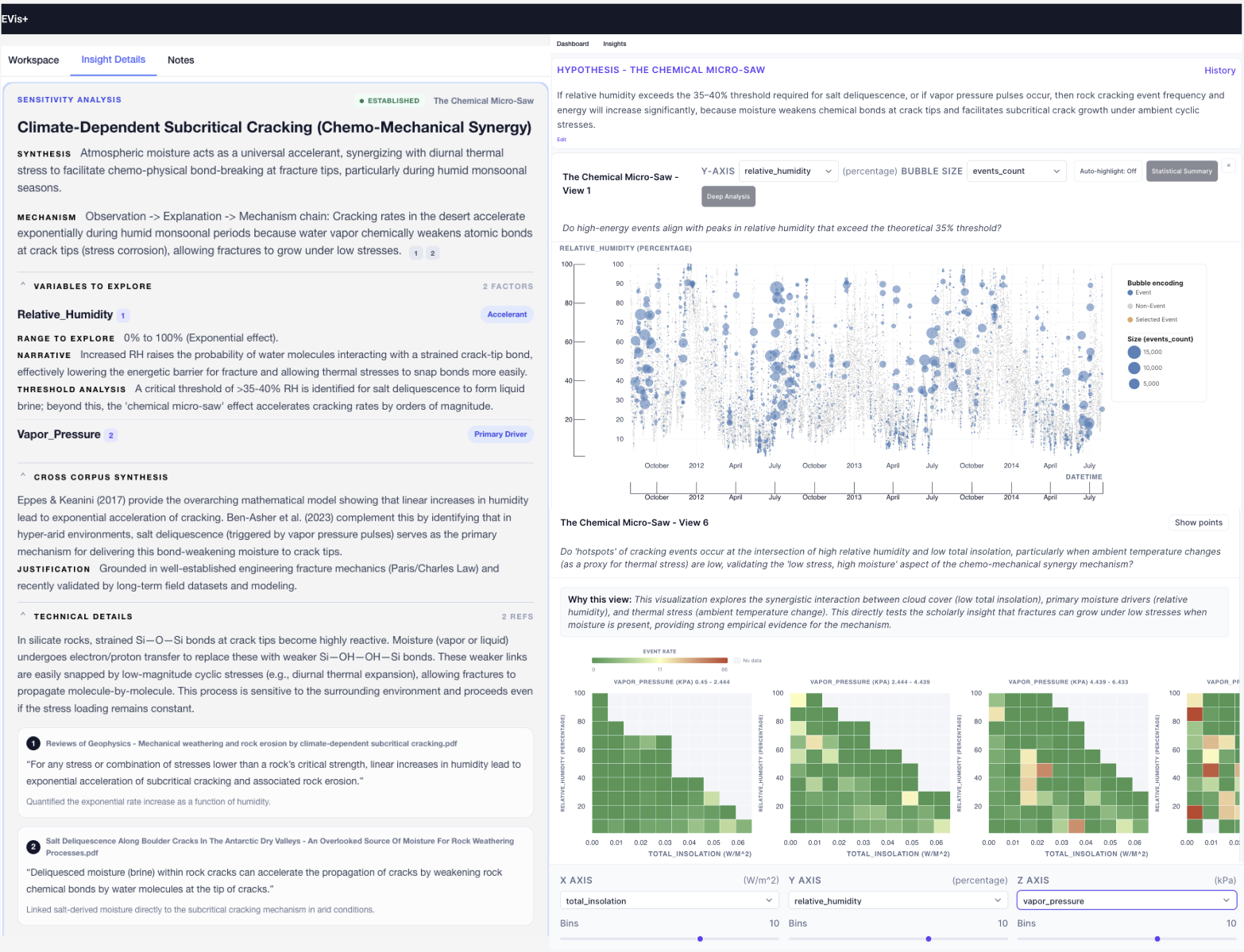}
  \caption{%
  Output for Case Study 2: Exploring the user-defined objective, “Analyze seasonal effects on rock cracking energy.” An LD-insight (left) guides the analysis
of seasonal cracking activity using coordinated temporal
and multivariate visualizations (right).
  }
  \label{fig:teaser}
}

\renewcommand{\manuscriptnotetxt}{}

\graphicspath{{figs/}{figures/}{pictures/}{images/}{./}} % where to search for the images

\usepackage{tabu}                      % only used for the table example
\usepackage{booktabs}                  % only used for the table example

\usepackage{mathptmx}                  % use matching math font

\AtBeginDocument{%
  \setlength{\abovecaptionskip}{3pt}%
  \setlength{\belowcaptionskip}{-4pt}%
}

\begin{document}

%%%%%%%%%%%%%%%%%%%%%%%%%%%%%%%%%%%%%%%%%%%%%%%%%%%%%%%%%%%%%%%%
%%%%%%%%%%%%%%%%%%%%%% START OF THE PAPER %%%%%%%%%%%%%%%%%%%%%%
%%%%%%%%%%%%%%%%%%%%%%%%%%%%%%%%%%%%%%%%%%%%%%%%%%%%%%%%%%%%%%%%

\firstsection{Introduction}

\maketitle

\enlargethispage{2\baselineskip}

\label{sec:intro} Exploratory analysis of observational data plays an important
role in scientific discovery. Researchers explore observational datasets
to uncover patterns, generate hypotheses, and develop mechanistic understanding
of complex phenomena. Such exploration rarely proceeds from data alone;
researchers discover and interpret patterns in light of prior knowledge and
empirical evidence \cite{Pirolli2005,Sacha2014}. Federico et al. argue that such
explicit knowledge should be integrated directly into data analysis systems
\cite{Federico2017role}. \added{In this paper, we focus on scientific literature as a
source of prior knowledge, complementing researchers’ domain knowledge and
experience. It provides findings that are grounded in published evidence,
contextualized by the conditions and settings in which they were observed, and
available for examination and reuse.}\revnote{R2a}

Through our long-term collaboration with geoscientists, we observed that
prior findings guide exploration toward relevant variables, ranges, and
thresholds; when anomalies arise, researchers consult the literature to assess
novelty, significance, and directions for further analysis.

\begin{figure}[tb] \centering \includegraphics[width=\columnwidth,
alt={}]{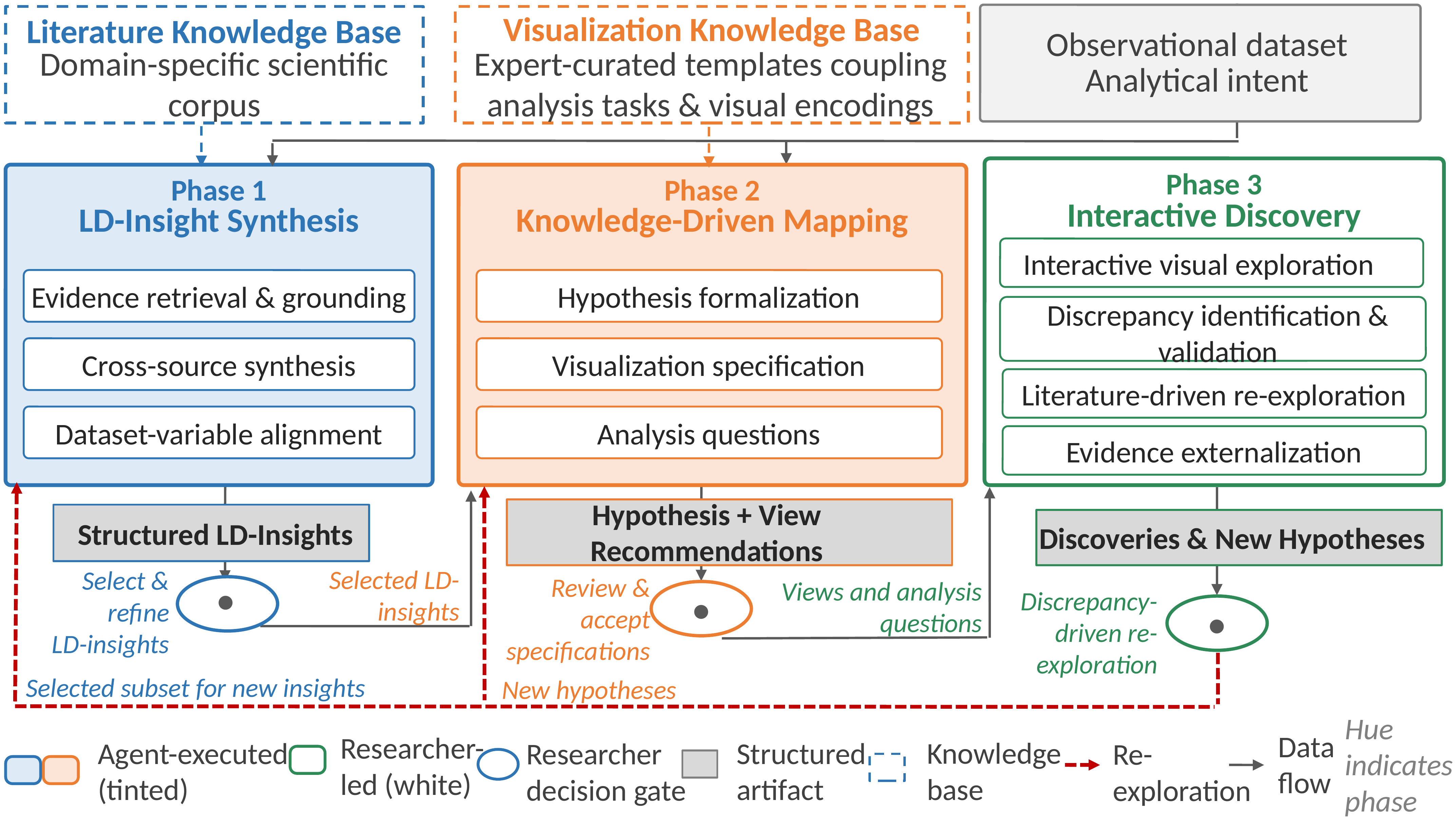}

\caption{The SciGroundVA workflow. Three phases each produce a structured
artifact, and the ovals are researcher decision gates: no artifact passes
forward until the researcher has reviewed and, where needed, revised it. The
dashed loop is discrepancy-driven re-exploration, which re-enters at Phase 1 for
new literature or at Phase 2 with the researcher's own hypothesis.}

\label{fig:workflow} \vspace{-15pt}

\end{figure}

In practice, however, this knowledge remains external to researchers'
analytical tools \cite{Brossier2026, Hutchinson2025}. While recent visual analytics systems
\cite{Shuai2025DeepVISBN,Weng2024,Zhao2024LightVALV,Zhao2024LEVAUL} have begun
to employ large language models (LLMs) for tasks such as intent inference, data
insight extraction, and visualization authoring, they largely focus on
information contained within the dataset itself and provide limited support for
leveraging literature-derived scientific knowledge during exploratory analysis.
Researchers therefore move back and forth between publications and
analytical tools, with extra effort to synthesize findings scattered across
paper collections and associate them with the dataset under analysis.
Transforming those findings into effective visual analyses further requires
substantial domain and visualization expertise, as researchers must determine
how they should be translated into analytical questions and investigated through
appropriate visualizations and interactions.

To address these challenges, we developed
EVis+, a literature-grounded visual analytics system for condition–event time
series analysis in geoscience (Fig. \ref{fig:workflow}). In such data,
continuously monitored environmental conditions are recorded alongside discrete
events characterized by their timing and magnitude. EVis+ is guided by
\textit{SciGroundVA}, a workflow that integrates
\textit{literature-derived insights} (LD-insights) synthesized from
scientific publications with domain-specific visual analytics techniques.
Each LD-insight is a structured representation of a reported finding and
the mechanism proposed to explain it, together with its source publications,
associated variables and conditions, and supporting evidence.
Researcher-selected LD-insights are formed into testable hypotheses, which in
turn determine the visualizations generated to examine them.\added{EVis+ builds on
recent mixed-initiative LLM-powered visual analytics systems
\cite{Zhao2024LEVAUL,Zhao2024LightVALV,Zhao2025ProactiveVAPV}, in which agents
propose analyses and researchers retain control, and extends them in two
respects. First, analyses are grounded in evidence extracted from a scientific
corpus rather than in the dataset alone, so each hypothesis carries the
empirical context under which prior findings were reported and can be traced
back to its sources. Second, agents instantiate expert-curated visualization
templates rather than synthesizing visualization code, keeping generated views
domain-appropriate and inspectable.}\revnote{R2b}

\deleted{In this paper, we propose SciGroundVA, a general visual analytics workflow to address these challenges.} \deleted{Although our implementation and evaluation are grounded in this domain, the workflow and its underlying design principles are domain-independent and could be adapted to other scientific contexts where researchers reason about observational data in light of prior empirical findings.}\revnote{R2c, SR: generality}

 This paper makes the following contributions:
\begin{itemize}[noitemsep,topsep=2pt,partopsep=0pt,parsep=0pt] \item
 \replaced{SciGroundVA, a literature-grounded
analytical workflow derived from long-term collaborations with geoscientists,
 showing how 
scientific  literature can be integrated into
observational data exploration and hypothesis-driven analysis.}{A visual analytics workflow, SciGroundVA, that couples agentic AI with expert-curated visualization templates to bring empirical knowledge from scientific literature into iterative, hypothesis-driven data exploration.}\revnote{R2b} 
\item
EVis+, a system instantiating this workflow for condition–event time
series analysis in geoscience, in which literature evidence remains traceable
from source excerpt through hypothesis to the visualizations that test it, and
AI-generated views are constrained to expert-curated templates rather than
synthesized code. \item \replaced{An evaluation combining an automated audit of
evidence grounding and attribution, a structured expert assessment of the
generated insights, and comparative and case studies with domain experts,
reporting both where literature-grounded analysis helped and where it fell
short.}{Case studies conducted on EVis+ with geoscientists analyzing observational datasets, illustrating how the workflow transforms the analytical process from ad hoc exploration into structured, mechanism-driven reasoning grounded in prior scientific knowledge.}\revnote{SR: evaluation} \end{itemize}

\section{Related Work}

\label{sec:related}

\subsection{Prior Knowledge and Sensemaking in Visual Analytics} Conceptual frameworks have modeled visual analytics as an
iterative process of foraging, hypothesis formation, and synthesis \cite{Pirolli2005,Klein2006,Sacha2014}, progressively
incorporating automated analysis \cite{Keim2008},
guidance and explicit knowledge integration \cite{Ceneda2017,Federico2017role}, semantic interaction
\cite{Endert2017}, and recently agent-based human–AI
collaboration \cite{Monadjemi2023}. A common thread is
that prior knowledge should function as a first-class component of the
analytical workflow. In practice, however, this knowledge largely remains
external to existing systems: design studies observe that domain knowledge and analytical reasoning typically
reside outside visualization tools, requiring
researchers to integrate external expertise with data
views \cite{Sedlmair2012}. Recent LLM-powered and
agent-based approaches \cite{ouyang2025nvagent,Wang2025Data,Weng2024} extend analytical
automation but, with a few exceptions, operate over the data at hand
without integrating empirically grounded scientific knowledge. The closest
system to ours, HypoChainer \cite{Jiang2026}, supports hypothesis-driven discovery through LLM
reasoning and biomedical knowledge graphs, but assumes an existing large-scale
knowledge graph and predictive model, focusing on filtering model predictions
rather than on open-ended exploration of
observational data.

\subsection{Literature-Driven Knowledge Extraction} Prior work on literature-driven knowledge extraction falls into two complementary
directions. Interactive visual analytics systems treat literature as a primary
dataset, enabling users to navigate research topics, relationships, and thematic
structures across corpora \cite{Hope2020,Lian2025,Wu2025}. In parallel,
NLP-based extraction pipelines automatically identify structured
information—claims, entities, and relations—from scientific texts using
pretrained language models \cite{Augenstein2019,Beltagy2019}, with recent
LLM-based approaches supporting richer extraction and organization
\cite{Agarwal2025,Wu2025}. \added{HypER \cite{Vasu2025HypER} extends this
direction to hypothesis generation, training a language model to validate
citation reasoning chains and generate hypotheses supported by them, preserving
provenance.}\revnote{new related work} However, both directions operate at the level of the literature
itself: extracted insights are not aligned with the variables, metadata, or
analytical scope of a specific observational dataset. Our work bridges this gap
by explicitly mapping LD-insights
to user-specified dataset variables and analytical intents, making them directly
actionable within visual exploration.

\subsection{LLM-Enhanced and Agentic Visual Analytics}
\added{LLM-powered visual analytics increasingly supports inspectable visualization generation and mixed-initiative analytical workflows. Across these systems, insights are typically derived from the dataset at hand and visualizations produced through open-ended generation. EVis+ changes both properties by grounding analysis in a scientific corpus and constraining generation to expert-curated templates.}\revnote{R2b}

\noindent\textbf{Stepwise, inspectable generation.} DeepVIS
\cite{Shuai2025DeepVISBN} embeds a five-stage Chain-of-Thought
 process into the NL2VIS pipeline, progressing from
chart type determination through data retrieval and refinement to final visualization,
 making each decision inspectable and correctable. We
 follow this stepwise principle but anchor the
chain in LD-insights rather than
natural language queries, and replace open-ended code generation with
 template instantiation, which
avoids the compilation failures DeepVIS reports
for specialized schemas.

\noindent\textbf{Mixed-initiative assistance.} LEVA \cite{Zhao2024LEVAUL}
uses the current system state and user selections to guide the LLM in choosing
relevant insight types. LightVA
 \cite{Zhao2024LightVALV} demonstrates a human-agent
collaboration model in which a planner recommends tasks,
the user selects and refines them, and an executor
generates visualizations and insights. EVis+ adopts this
mixed-initiative posture but changes what drives it: recommendations are
anchored in prior knowledge rather than data-driven task suggestions, and a
researcher-selected data subset triggers retrieval of further literature
evidence rather than data-internal computation.

\subsection{Visualization Templates and Intent-Oriented Design} Reusable
visualization templates, such as those produced by Harper and Agrawala
\cite{Harper2018} and DataWink \cite{Xie2026DataWink}, are created by extracting
styles or encodings from existing charts; most capture graphical structure
without modeling the analytical purpose a visualization serves. Visualization authoring systems such
as Drillboards \cite{Shin2025Drillboards} and Data Formulator 2
\cite{Wang2025Data} focus on lowering the barrier to creating expressive
visualizations without explicitly modeling analytical intent.
 Prior work on
visualization recommendation has explored how user tasks, data characteristics,
and perceptual guidelines can inform visualization choices
\cite{Luo2021,Mackinlay2007}. Several systems model analytical intent
explicitly: TaskVis \cite{shen2021taskvis} pairs 18 classical analytic tasks
with suitable chart types; MEDLEY \cite{Pandey2023MEDLEY} recommends dashboards
from four generic data-level intents; LightVA \cite{Zhao2024LightVALV}
translates user goals into typed tasks such as trend analysis and outlier
detection; LEVA \cite{Zhao2024LEVAUL} maps the current task and interaction
state to insight types; and Waltzboard \cite{Choi2024Waltzboard} has users
weight five effectiveness measures such as specificity and diversity to optimize
dashboard composition. In each case the intents are predefined generic
categories or abstract data-level measures. In contrast, our workflow
derives analytical intents from user-selected
LD-insights and uses them to select and rank domain-specific
visualization templates.

\section{Design Requirement}

\label{sec:design-requirement} \addedhere{terminology: \emph{analyst} is now \emph{researcher} throughout, since every collaborator and participant is a practising scientist working in their own research area}\revnote{R2a: audience}

\added{We developed design requirements through
close collaboration with a geoscientist co-author (E1, 24 years on rock
cracking), iterative prototyping, and formative interviews. Observing E1's
analytical workflow with EVis~\cite{Feng2022EVis}, our prior system for
multivariate time-series analysis, we found their exploration guided by prior
knowledge that resided in their expertise rather than in the analytical
environment, and that drawing on it required manual effort, visualization and
domain expertise, and familiarity with the literature that less experienced
researchers may lack. To validate and refine the resulting requirements, we
conducted 60-minute remote interviews with geoscientists E2 (27 years, water
turbulence) and E3 (3 years, sedimentology). Each supplied a paper collection,
manually extracted insights from it as a reference, then assessed the
LD-insights our prototype produced and discussed how they would use them in
their own analysis. Their feedback confirmed and extended the requirements
identified with E1.}\revnote{R3.1, R2c: methodology}

\deleted{We developed the design requirements through a year-long design study combining close collaboration with a geoscientist co-author (E1), iterative prototyping, and formative interviews with additional geoscientists. The process unfolded in three overlapping stages.}

\noindent\textbf{R1: Credible literature sources and semantic alignment}.
To make LD-insights actionable, the system should draw them from sources the
researcher can verify and trust, bridge those findings with the quantitative
schema of the researcher's dataset, and shape them around the researcher's
current analytical intent.

\added{\textbf{R1.1: Source reliability}. E1 repeatedly emphasized that the
corpus should consist of publications the researcher considers scientifically
credible. Corpus composition should remain under researcher control.}\revnote{R3.2: corpus construction}

\textbf{R1.2: Source evidence grounding}. Extracted insights should
reflect a paper's own empirical findings rather than claims it merely cites, and
researchers should be able to verify them by tracing each insight to its source
excerpt and surrounding context. Early prototyping showed that when this
distinction was absent, E1 quickly lost trust in the output.

 \textbf{R1.3:} \textbf{Entity mapping}.
 Researchers should be able to examine how variables,
causal factors, and experimental conditions in the
literature map to dataset variables
and metadata. The system
should automatically generate these mappings while allowing researchers to verify and correct them,
 supporting assessment of whether and
under what conditions reported findings apply to the data.

\added{\textbf{R1.4: Analytical-intent alignment}. The system should use the
researcher's analytical intent to guide literature retrieval and LD-insight
synthesis, focusing the resulting insights on the relationships under
investigation rather than producing generic corpus-level summaries.}\revnote{R3.3}

\noindent\textbf{R2: LD-insight synthesis and exploration}. Our early
experiments with embedding-based retrieval approaches (SciBERT,
Sentence-Transformers) generated a large number of insights in fragmented text
chunks, making it difficult for E1 to identify actionable findings. E2
noted that relevant evidence is often distributed across studies, with each
contributing only partial evidence. The system should therefore construct
LD-insights across the corpus and support exploration of the resulting insight
space.

\added{\textbf{R2.1: Cross-corpus insight synthesis and evidence trajectory}.
The system should synthesize related evidence across publications into mechanism-level LD-insights while preserving conflicting findings and tracing how evidence develops across sources. E2 and E3 emphasized that such a cross-source view helps researchers calibrate confidence in literature-derived expectations, identify evidentiary gaps, and judge whether their own observations align with prior evidence or may be novel.}\revnote{R3.3}

\textbf{R2.2: Multi-resolution LD-insight exploration}. Researchers
should be able to browse and search a large collection of LD-insights and
examine insights of interest at greater levels of detail. E3 noted that
insufficient technical detail for complex, multi-variable phenomena limited the
usefulness of high-level insights in nuanced analysis.

\noindent\textbf{R3:} \textbf{Enable 
\replaced{hypothesis-driven}{expectation-driven} visual analysis}.\revnote{R4d} E1's workflow showed that
expert analysis is guided by expectations grounded in prior knowledge, reasoning
they performed mentally. Translating prior knowledge into visual analysis
requires formulating a testable hypothesis, selecting visualizations that can
reveal relevant patterns, and assessing whether those patterns support or
contradict it. The system should externalize this reasoning, using visualization
to examine expectations \cite{Hullman2021} and mapping analysis questions to
visual representations \cite{Brehmer2013,Munzner2014}.

\textbf{R3.1: Externalized reasoning chain}. The system should
externalize the reasoning chain from LD-insight to hypothesis, visualization
specifications, and analysis questions so that researchers can inspect each step
and judge whether the recommendation aligns with their expectations and
analytical goals.

\added{\textbf{R3.2: Domain-appropriate visual representations}. Recommended
visualizations should reflect analytical practices established in the domain
rather than generic chart choices. E1's motivation for the collaboration was to
obtain views tailored to their research problem.}\revnote{R2b}

\textbf{R3.3: Interactive reasoning control}. Researchers
should be able to revise the intermediate reasoning artifacts, including
hypotheses, visualization specifications, and analysis questions, rather than
treating system-generated recommendations as fixed. E1's iterative refinement in
EVis demonstrated that researchers' reasoning evolves during exploration, so
rigid outputs would disrupt rather than support it.

\noindent\textbf{R4: Support for exploratory expansion}. Discrepancies
between observed patterns and expectations should be treated as productive
signals for exploration rather than errors \cite{Hullman2021} and as triggers
for reframing during sensemaking \cite{Pirolli2005,Sacha2014}. When E1 found that low ambient
temperature alone did not explain all large cracking events,
they expanded the analysis to additional variables, a
process the system should support systematically.

\textbf{R4.1: Discrepancy validation}. Researchers should be able to
examine potentially anomalous subsets using summary statistics and comparative
analysis to assess whether they reflect genuine discrepancies or artifacts of
sample size, data coverage, or measurement conditions, supporting more informed
decisions about whether to pursue a new analytical direction.

 \textbf{R4.2: Literature-driven re-exploration}. When a
validated discrepancy cannot be explained by the current hypothesis,
 researchers should be able to retrieve additional
LD-insights relevant to the observed pattern and formulate
new hypotheses, initiating a new cycle of hypothesis-driven analysis.
New exploration should build on the existing analytical state, with prior
views providing context.

\section{The SciGroundVA Workflow}

\label{sec:workflow} Building on these requirements, we propose SciGroundVA, a
\replaced{visual analytics workflow presented as a design abstraction instantiated
within EVis+. The high-level analytical process and design principles reflect
lessons learned from our design study, whereas the specific corpus, visual
templates, and condition–event mapping reflect the domain constraints of our
geoscience deployment.}{While we instantiate and evaluate the workflow for a specific scientific context in Section 5, the underlying architectural choices are designed to generalize to other domains where researchers reason about observational data in light of prior empirical knowledge.}\revnote{R2c, SR: generality}

\subsection{Overview} SciGroundVA (Fig.~\ref{fig:workflow}) relies on two
knowledge bases (KBs) that must be constructed before analysis begins: a
literature KB from which LD-insights are extracted and synthesized, and a
visualization KB of expert-curated templates. The workflow has three phases, and
human judgment gates every transition. Phase 1 extracts and synthesizes
LD-insights from the literature KB, and the researcher selects which to pursue.
Phase 2 turns a selected LD-insight into a hypothesis and a set of views
instantiated from the visualization KB, which the researcher accepts or revises
before they are built. Phase 3 is interactive exploration of the dataset through
those views, where a discrepancy the researcher judges genuine can start a new
cycle of retrieval and analysis.

\subsection{Dual Knowledge Bases} The literature KB consists of domain-specific
scientific publications relevant to the research problem. Full text is
essential because evidence grounding requires direct access to empirical
observations, experimental setups, and methodological contexts. To ensure the
quality of the LD-insights, this KB should include only publications specified
by the researcher, supplied as PDF files, as a paper list, or through search
terms over a reliable publication source (R1.1). The visualization KB contains
domain-specific visualization templates co-designed by visualization experts and
domain experts, ensuring that instantiated views reflect established analytical
practice in the domain (R3.2). Each template defines an analytical purpose,
compatible data roles and visual channels, configurable parameters, and
supported interactions. \added{Together, these knowledge bases place boundaries on the
role of automation. The literature KB constrains what evidence the system can
retrieve and use, while the visualization KB constrains how that evidence can be
explored and analyzed. This separation allows a system to leverage LLMs for
flexible, knowledge-driven reasoning without giving them unrestricted control
over either the evidence base or the visualization design space.}\revnote{R1, R3.2}

\subsection{Phase 1: LD-Insight Synthesis} This phase transforms empirical
observations from the literature into structured LD-insights in three
steps: retrieving those relevant to the researcher’s analytical intent, grouping
related observations across sources into mechanism-level representations, and
aligning those mechanisms with the researcher’s dataset variables (R1, R2).

\noindent\textbf{Evidence retrieval and grounding.} Retrieval operates
over the literature KB (R1.1) and is conditioned on the researcher’s analytical
intent and available dataset variables, so the results are focused on the
relationships under investigation (R1.4). It extracts empirical observations
reported by the studies rather than claims appearing only as related work or
cited background. This constraint ensures that downstream
reasoning is anchored in verifiable primary evidence, reducing the risk of
propagating unsupported claims (R1.2). Each extracted observation preserves
explicit provenance through source attribution and contextual excerpts, allowing
 researchers to trace any synthesized claim back to
its origin (R1.2).

\noindent\textbf{Cross-source synthesis.} Observations describing the same phenomenon are grouped across the corpus, and a single mechanism is proposed for each group. Each mechanism is assigned a scientific standing: {\it well-established} where multiple independent sources corroborate it, {\it emerging} where support is limited, and {\it anomalous} where an observation may contradict prevailing evidence. The synthesis records how consistently the corpus supports the mechanism and traces how different sources complement each other: for example, one study may provide initial field evidence, another may formalize the underlying explanatory model, and a third may identify additional triggering factors. This cross-source narrative enables researchers to assess not only what is known about a mechanism but how that understanding was built and where evidentiary gaps remain (R2.1).

\noindent\textbf{Dataset-variable alignment.} The synthesis also proposes
literature variables that may correspond to dimensions in the researcher’s
dataset, their roles in the mechanism, an explanation of how each influences it,
and, where available, relevant value ranges, thresholds, or qualitative
conditions. Researchers can inspect the proposed alignment together with its
supporting evidence to judge whether terminology, measurement definitions,
units, experimental conditions, and other assumptions are sufficiently
compatible with their observational data (R1.3). The three steps together produce an LD-insight, presented at
multiple levels of detail, from concise
summaries to detailed breakdowns of
variables, conditions, and source evidence, supporting
multi-resolution exploration of the insight space (R2.2). At the transition
between Phase 1 and Phase 2, the researcher searches,
browses, and examines the synthesized insights and selects those most
interesting to them. This gate ensures that only insights the
 researcher judges credible
drive the hypothesis formalization and visualization design
that follow.

\subsection{Phase 2: Knowledge-Driven Mapping} This phase translates structured
LD-insights into analytical specifications and coordinated visualizations
(R3). It externalizes the reasoning as a hypothesis, visualization
specifications, and analysis questions, produced in three steps: formalizing a
testable hypothesis, specifying the visualizations that would confirm or
contradict it, and attaching to each view the question the researcher should ask
when reading it.

\noindent\textbf{Hypothesis formalization.} A formalized hypothesis is
proposed for the selected LD-insight, encoding the triggering conditions, the
expected outcome, and the mechanism that connects them. The hypothesis is paired
with a core visual question naming the relationship that must be seen to
evaluate it. Researchers can inspect and revise both before proceeding (R3.1).

\noindent\textbf{Visualization specification.} The core visual question
is matched against the analytical task annotations in the visualization KB to
select templates (R3.2). Where the hypothesis names several drivers, separate
templates are selected for each driver. Each template is then specified by
binding the dataset variables identified during alignment to its visual
channels, together with the range of conditions under which the effect should
appear and a contrasting range where it should not.

\added{\noindent\textbf{Analysis questions.} Each view is paired with an
analysis question directing the researcher’s attention to the pattern it is
meant to reveal. These questions state what would confirm or contradict the
hypothesis before the researcher examines the data, so each view is read as
evidence rather than as a general display.}\revnote{R4d: hypothesis vs.\ analysis question} Researchers inspect the hypothesis,
the visualization specifications, and the analysis questions, and revise any of
them before accepting. On acceptance, the specified views are instantiated
against the dataset.

\subsection{Phase 3: Interactive Discovery} The interactive discovery phase
supports researchers in examining whether the observational data are
consistent with, complicate, or contradict the expectations represented by the
current hypothesis.

\noindent\textbf{Interactive visual exploration.} The visualizations
instantiated in Phase 2 are presented as coordinated views, with the supporting
and contrasting condition ranges marked. The hypothesis and its analysis
questions remain visible alongside the views. Researchers can select a subset of
data items and see it highlighted across all views, comparing its behavior
against different drivers, and can examine summary statistics for that subset
for quantitative context. Researchers retain full control over the dashboard and
can reconfigure any view as their interest evolves (R3.3).

\noindent\textbf{Discrepancy} identification and validation. When observed
patterns diverge from expectations, researchers can
select the anomalous subset and examine its summary
statistics, comparing them against
those of other subsets they select, to assess whether the discrepancy is
genuine before committing to a new analytical direction (R4.1).

\noindent\textbf{Literature-driven} re-exploration. When a validated
discrepancy cannot be explained by the current hypothesis, the
researcher can trigger re-exploration: additional LD-insights relevant to the observed pattern are retrieved,
synthesized into new mechanisms, and used to formulate new hypotheses with
associated visualization specifications (R4.2). This re-enters the
workflow at Phase 1, with the resulting views added to the existing dashboard. \added{A
discrepancy may also prompt a hypothesis of the researcher’s own: they can enter
it directly and re-enter at Phase 2, or use it as a new analytical intent and
re-enter at Phase 1, retrieving literature against which the novelty and
significance of the observation can be assessed.}\revnote{R4}

\noindent\textbf{Evidence externalization.} Researchers save discoveries
as notes tied to the hypothesis under investigation. The analytical products of
the workflow are thus hypotheses linked to their scholarly origins and supported
by researcher-curated evidence, together with discrepancies that prior findings
do not account for and that can prompt the researcher's own new hypotheses.
\added{Throughout the workflow, SciGroundVA does not treat LLMs as general-purpose
reasoners trusted to extract scientific knowledge or to determine
correspondences between literature and data on their own; they perform bounded
transformations within a structured workflow, leaving direction and
interpretation to the researcher.}\revnote{R3.2, R3.3, R1}

\section{EVis+}

\label{sec:prototype} We propose EVis+, an agentic visual analytics system for
condition-event time series analysis in geoscience, built on our prior
EVis system~\cite{Feng2022EVis}, and the system within which the workflow of
Section~\ref{sec:workflow} is instantiated. \added{It coordinates two specialized
agents: a Scholar Agent for LD-insight synthesis (Phase 1), and an Analysis
Agent that turns a selected LD-insight into a testable hypothesis and the
coordinated views that examine it (Phase 2). Phase 3 introduces no agent;
exploration is researcher-driven.}\revnote{R4b: agent definition}

\subsection{Knowledge Bases and Dataset Context}

\label{sec:kb-construction} EVis+ draws on three resources throughout the
workflow. The researcher supplies a literature corpus alongside the
observational dataset for each analysis (R1.1); the visualization knowledge base is
built into the system and reused across condition-event time series datasets.
Together, they provide the context in which the agents interpret literature and
formulate visual analyses.

\noindent\textbf{Literature knowledge base.} The corpus of full-text
publications relevant to the dataset is uploaded to an OpenAI vector store and
indexed for retrieval, and is used across analysis sessions on that dataset.

\noindent\textbf{Visualization knowledge base.} We constructed a library of
visualization templates tailored to condition-event time series analysis.
 Templates encode decisions such as how to
 overlay event metrics on condition variables,
informed by sustained collaboration with geoscientists (R3.2).
Each fills in the schema fields introduced in Section~\ref{sec:workflow}:
analytical purposes such as examining diurnal patterns or comparing event rates
across conditions; channels such as x-axis, y-axis, color, size, and faceting;
parameters such as temporal resolution, binning strategy, and event metric; and
interactions such as brushing, linking, and filtering. \added{The current library
contains six templates: time series scatterplots, diurnal scatterplots, standard
scatterplots, faceted heatmaps, histogram small multiples, and layered temporal
bubble matrices, building on visualizations previously developed and validated
with domain experts in EVis~\cite{Feng2022EVis}.}\revnote{R2b: template library}

\added{\noindent\textbf{Dataset context.} Before analysis, EVis+ preprocesses
the target dataset into a compact schema used as context for the agents. For
each variable, it records the name, inferred type, and, when applicable, range,
standard deviation, counts of unique and missing values, and examples. Three
sample rows preserve values in their original form, while event variables are
identified separately to distinguish environmental conditions from event
occurrence and magnitude. This schema provides a shared vocabulary for mapping
literature concepts to dataset fields, and most agent reasoning operates over it
rather than the full dataset. During discrepancy-driven re-exploration, selected
data items are added as raw rows restricted to variables encoded in the
originating view, together with the current LD-insight, hypothesis, and
dashboard.}\revnote{R4c: selected data as analytical context}

\subsection{LLM Agent Configuration}

\label{sec:agents} \added{The two agents are sequenced by application code rather than
by a planning agent; each performs a bounded transformation and produces
a structured intermediate representation passed to the next stage. The
implementation uses the OpenAI Responses API with GPT-5-mini for literature
synthesis, hypothesis generation, and visualization recommendation. Before
literature retrieval, a lightweight GPT-5-nano call checks whether the
researcher's analytical intent specifies a sufficiently concrete variable,
relationship, or phenomenon; if the intent is underspecified, the system prompts
the researcher for clarification rather than proceeding with retrieval. Agent
outputs are generated directly into predefined Pydantic schemas, ensuring
consistent structure and eliminating the need to parse unconstrained
natural-language responses.}\revnote{R4b, R1: implementation detail}

\noindent\textbf{Scholar Agent.} Given the researcher's analytical intent
and dataset variables, this agent retrieves evidence from the indexed corpus and
organizes it into structured, mechanism-oriented LD-insights. Each captures a
mechanism summary and explanation, associated variables and roles, ranges or
thresholds, a suggested analytical task such as trend or sensitivity analysis,
cross-paper synthesis, and supporting source evidence. Variable roles use a
constrained set of Primary Driver, Accelerant, Threshold Trigger, Modulating
Factor, and Inhibitor, while source-evidence records preserve the quoted
passage, the title of its source paper, and its contribution to the synthesis.
The agent also assigns descriptive scientific standing labels based on
cross-source agreement, established, emerging, or anomalous, to organize the
insight space; these are not calibrated measures of scientific confidence.

\added{When required fields are missing from a returned insight, the system
issues a single follow-up call that supplies the partial insight and asks the
agent to fill only the missing fields. Insights that remain incomplete after
this call are retained and logged rather than discarded or regenerated. We
developed the extraction prompt iteratively, using manually identified findings
as a reference. One annotator annotated a 31-paper corpus for findings
associated with 11 dataset attributes, and a second examined the results,
yielding 245 findings. We supplied these together with the 211 insights produced
by an early version of the agent to an LLM, and asked it to characterize the
properties the manually identified findings had that the agent's output lacked.
Chief among these was that findings should be traced to a paper's own empirical
results rather than to claims it cites, which is encoded as the
evidence-filtering constraint in the current prompt (R1.2).}\revnote{R3.2, R3.3: evidence that the extraction step works}

\noindent\textbf{Analysis Agent.} Given a researcher-selected LD-insight and the dataset context, the Analysis Agent implements Phase 2 of SciGroundVA by generating a testable hypothesis and recommended views. Each view specifies its analysis question, variable encodings, expected pattern, and relevant supporting or contrasting conditions. Recommendations are validated against the dataset schema before rendering; editing the hypothesis regenerates the recommendations.

\added{Recommendations are checked before rendering to confirm that referenced
variables exist and required data roles are satisfied, and are returned for
targeted revision if they fail.}\revnote{R1: reliability}

\subsection{User Interface}

\label{sec:ui}

\subsubsection{Design Principle: Externalize the Reasoning Process.}

\deleted{A central design decision in the EVis+ interface is the use of hypotheses as the central organizing element.}\revnote{R4d}
EVis+ makes the progression from literature to hypothesis to evidence explicit and revisable. The hypothesis serves as the organizing unit connecting the selected LD-insight, the views used to examine it, and the notes recording the researcher's observations, and remains inspectable and revisable throughout the analysis (R3.1, R3.3). Each view is paired with an analysis question that directs attention to the pattern it is intended to examine.

The dashboard accumulates rather than resets as the analysis progresses. New views are added at the top when the researcher pursues a new hypothesis or interactively adds a view to the current analysis. These views are coordinated with existing ones, supporting continuous, progressive visual exploration (R4.2).

\subsubsection{Views} A left sidebar holds the Workspace, Insight Detail, and
Notes tabs; the main area switches between
the Insights view and the Dashboard.

\noindent\textbf{Workspace.} The Workspace tab manages datasets, event attribute
configuration, analytical intents, hypotheses, and literature corpus. A
companion preprocessing utility converts datasets in common tabular formats to
an EVis+-compatible format.

\noindent\textbf{Insights.} The Insights view on the right presents the
structured LD-insights produced by the Scholar
Agent as interactive cards that can be browsed, searched, and filtered.
 Each card presents an LD-insight at multiple levels of
detail (see Fig.~\ref{fig:teaser} for a card in full detail): a short
mechanism alias for easy reference, the
analytical task, a concise summary, and the
mechanism. Expandable sections reveal associated variables with their roles and
value ranges, cross-corpus synthesis, and technical details, which are the
underlying source evidence with source excerpts from the original papers,
annotated with their specific contributions to the synthesized mechanism, as
well as variable mappings to the researcher's
dataset schema. This supports the provenance verification
(R1.2). Researchers can launch an integrated PDF
reader (displayed as a pop-up window) to inspect the original document by
clicking the numbered circle in the card. When researchers find an
LD-insight of interest, they click the Recommend Visualization button at the end
of its card. EVis+ sends the LD-insight to the Analysis Agent, which returns an
editable hypothesis together with the recommended views and their analysis
questions. Researchers can edit the hypothesis, the views and their
configurations, and the analysis questions, or add additional views, and approve
the recommendation to instantiate the views in the Dashboard.

\noindent\textbf{Insight Detail.} When an LD-insight is under exploration, the Insight Detail
tab on the left sidebar displays its card in full
 detail, serving as a persistent reference that
 researchers can consult while working in the
Dashboard without switching away from the visualization.

\noindent\textbf{Dashboard.} The Dashboard presents the coordinated
visualizations generated by the Analysis Agent.
Each visualization is annotated with a targeted analysis question derived from
the hypothesis, guiding the researcher's attention
to specific patterns. The supporting and contrasting condition ranges are
shaded so that the regions where the hypothesis predicts the effect can be
compared against those where it does not, and the shading can be toggled off.
Views are linked through brushing and highlighting, enabling cross-view pattern
examination.

\noindent\textbf{Notes.} The Notes tab allows researchers to save discoveries as structured note cards tied to the hypothesis under investigation, capturing selected data subsets, views, user comments, and summary statistics. Clicking a note restores the corresponding selection for re-examination.

\noindent\textbf{Statistics.} The statistics view displays event rates for a
user-selected data subset; researchers compare subsets by selecting each
in turn, supporting discrepancy validation (R4.1).

\subsubsection{Interactive, Progressive Visual Exploration}

\noindent\textbf{Researcher-driven workflow steering.} Researchers can revise intents, LD-insights, hypotheses, visualization recommendations, and view configurations throughout the workflow, while retaining final control over all analytical decisions (R3.3).

\noindent\textbf{Evidence externalization.} Notes externalize the evidence accumulated during exploration and can be compared side by side or exported with the associated LD-insight and hypothesis, producing a traceable record from literature-derived knowledge to hypothesis and observed evidence.

\noindent\textbf{Deep analysis and re-exploration.} When observed patterns
diverge from expectations, researchers can select the
anomalous subset and trigger deep analysis. \added{The system first
performs an alignment audit, checking the selected data items against the
existing LD-insights and classifying them as \textit{Aligned}, naming the
insight that accounts for them, or \textit{Contradictory}. Only when no existing
insight accounts for them are additional LD-insights retrieved, searching on the
conditions present in the selected data items rather than the original
analytical intent; the results are appended to the session's LD-insight set and
remain open to the same inspection as the initial ones.}\revnote{R4c} The researcher remains
in the Dashboard throughout: the LD-insight that accounts for the selection,
whether matched or newly retrieved, becomes the active LD-insight, complementary
views focused on the selected data items are added to the existing dashboard,
and the selected items stay highlighted across it, so that each exploration
cycle builds on rather than replaces the researcher's accumulated analytical
state (R4.2).

\section{ Evaluation}

\label{sec:case-study} \addedhere{section retitled from \emph{Case Study}; three of the four studies below are new}\revnote{SR: evaluation}

\added{We evaluated EVis+ through four complementary studies: a grounding audit, a researcher assessment of the generated LD-insights, two case studies, and a comparative study. Eight researchers, including two co-authors, participated across these evaluations. One participated in two of them. The evaluation used eight corpora spanning three disciplines. The corpora were: in geoscience, C-Geo (geomorphology, 31 papers), C-HydGeo (hydrogeology, 23), C-HydSed (hydroacoustic sediment transport, 6), and C-Orb (orbital cyclostratigraphy, 14); in architecture, C-Mic (microalgae biomass, 10); and in computer science, C-ML (neurological disorder detection, 11), C-Sec (software security, 25), and C-Pri (data privacy, 41).}

\subsection{Evidence Grounding and Attribution Evaluation}

\label{sec:grounding} \added{We evaluate whether LD-insights remain traceably connected
to primary evidence (R1.2): whether quoted passages exist within the
corpus, and whether grounded claims correctly credit their source publications.}\revnote{R3.2, R3.3, SR: evidence for LLM steps}

\noindent

\added{\textbf{Method:} We audited all 1,338 quoted passages from the 394 LD-insights generated across the eight corpora. For each passage, an automated verifier searched the corresponding corpus for the best-matching sentence window, scoring similarity as a normalized edit distance after normalizing case, punctuation, and hyphenation, so that only differences in wording affect the score. Passages were classified as verbatim, near-verbatim, paraphrased, or ungrounded. For grounded passages, we also compared the publication the system cited against the one containing the best match.}

\noindent

\added{\textbf{Results:} Pooled across the eight corpora, 92.7\% of the 1,338 quoted passages could be
grounded in the literature: 24.1\% verbatim (exact substring), 41.9\%
near-verbatim (at least 90\% similarity), and 26.8\% paraphrased (60--90\%
similarity). Among the 1,240 grounded passages, 95.4\% were attributed to the
correct publication. Per corpus, as passages, grounded, and attributed:
C-Pri (638, 93.1\%, 97.6\%), C-Geo (198, 88.9\%, 93.8\%),
C-HydGeo (168, 95.8\%, 82.6\%), C-Orb (111, 96.4\%, 99.1\%),
C-Mic (80, 76.2\%, 95.1\%), C-HydSed (51, 100\%, 100\%),
C-Sec (48, 95.8\%, 100\%), and C-ML (44, 100\%, 100\%).
Grounding was lowest in C-Mic and attribution in C-HydGeo, which together with
C-Geo accounts for 39 of the 57 attribution errors. C-Geo, the corpus used in
developing the extraction prompt (Section~\ref{sec:agents}), grounded at 88.9\%,
below the pooled rate, indicating that the extraction constraints are not tuned
to the corpus from which they were derived.}

\subsection{Researcher Assessment of LD-Insights}

\begin{figure}[t] \centering
\includegraphics[width=\columnwidth]{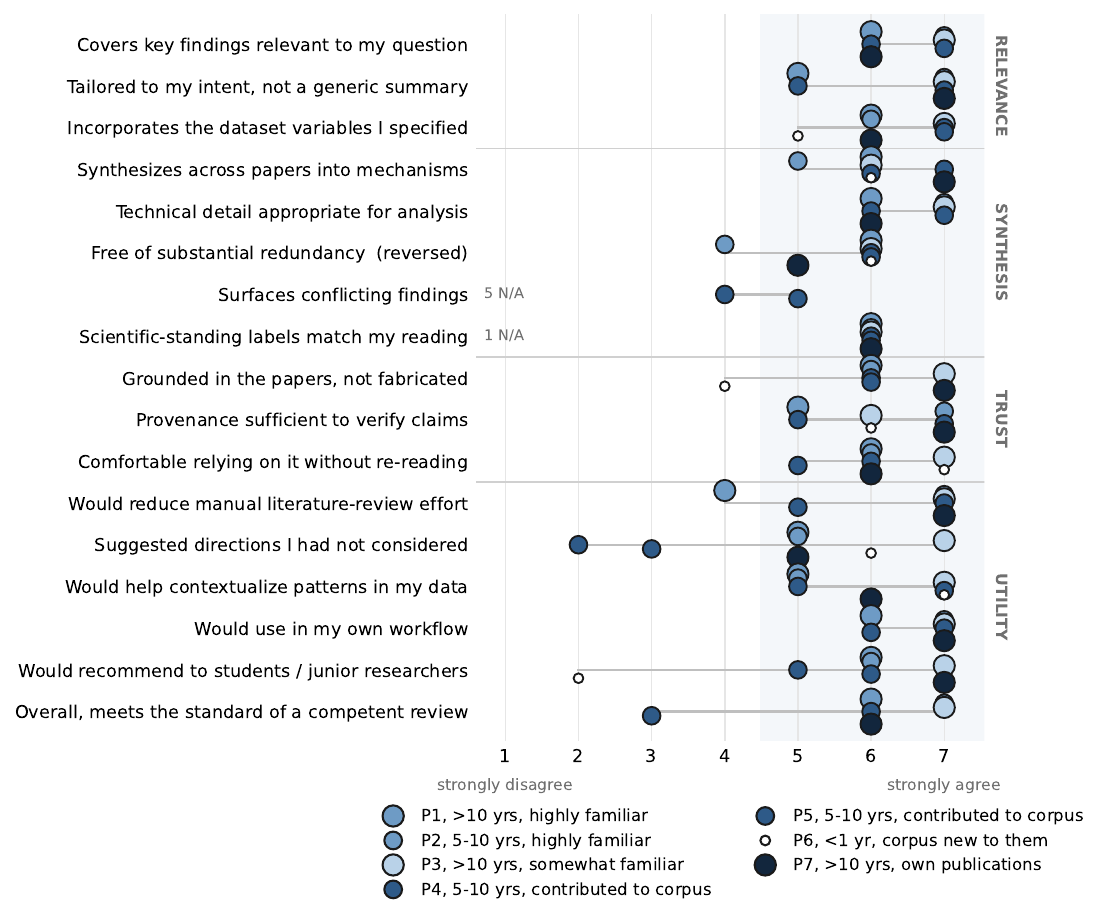}
\caption{Per-participant ratings of the generated LD-insight set. Each row is
one questionnaire statement; each marker is one participant, dodged vertically
so equal ratings remain visible. Marker area is proportional to years of
research experience and color encodes the participant's prior relationship to
the corpus. The shaded region marks agreement.}

\label{fig:ratings} \vspace{-15pt}

\end{figure}

\added{We conducted a structured assessment with seven researchers to evaluate
LD-insight sets extracted from their research areas for their relevance,
synthesis, trust, and utility (Fig.~\ref{fig:ratings}).}\revnote{SR: evaluation is limited}

\added{\noindent\textbf{Method.} Participants P1--P7, one of whom is a co-author, ranged from less than one year to more than ten years of research experience. Six self-identified as experts in the research area represented by the corpus they assessed, while one identified as a novice. Each expert assessed a corpus from their own research area: two had contributed papers to the corpus, one had authored or co-authored all papers in it, two were highly familiar with its publications, and one recognized some of them. The novice participant, P6, was instead assigned C-Mic, a corpus new to them that was also assessed by an expert participant.}

\added{The assessment covered six corpora across three disciplines: C-HydSed, C-HydGeo, and C-Orb in geoscience; C-Mic in architecture; and C-ML and C-Sec in computer science. None of these corpora was used in developing the extraction prompt. For each participant, we generated an LD-insight set based on their analytical intent and assigned corpus. Participants received the LD-insight sets as screenshots by email, with each card expanded to show full details, including source excerpts and variable mappings, and submitted their ratings through an online form. The resulting judgments therefore assess LD-insight quality independently of the interactive EVis+ interface.}

\added{Participants rated seventeen statements on a seven-point agreement scale
(Fig.~\ref{fig:ratings}). On the item concerning conflicting findings,
participants could answer not applicable if they were unaware of substantive
disagreements in their corpus, and five did so. Participants also answered
free-text questions on expected time savings, features that would increase their
confidence, anticipated difficulties in their own domain, and reasons for any
low rating. Ratings are reported per participant (Fig.~\ref{fig:ratings}); with
seven respondents we make no distributional claims.}

\added{\noindent\textbf{Results.} \emph{Relevance} was rated uniformly high:
every participant agreed the set covered the key findings relevant to their
question, was tailored to their stated intent rather than reading as a generic
corpus summary, and incorporated the dataset variables they had specified
(R1.3, R1.4).}

\added{On \emph{synthesis}, all seven agreed the set aggregated findings across
papers into mechanisms rather than restating individual papers, and that the
level of technical detail suited subsequent analysis. All six who judged the
scientific-standing labels rated them consistent with their own reading of the
literature. No participant judged the set to contain substantial redundancy,
although this item was reverse-worded and four participants revised their answer
to it on review.}

\added{\emph{Trust} was high throughout: of the twenty-one trust ratings, only
one fell below 5, P6's 4 on whether the insights were grounded rather than
fabricated. Two participants identified failures of different kinds, and
together they separate evidence grounding from scientific correctness. P6 found
an insight stating that carbon dioxide above ten percent reduces algal growth
where the cited paper had tested only the ten percent level, an
overgeneralization beyond the source that the grounding audit cannot detect
because the quoted passage itself is present in the corpus. P2 conversely noted
an insight faithfully reporting a claim of ninety-nine percent classification
accuracy that they judged implausible, observing that a correctly extracted
insight may still carry an unreliable claim from its source. Provenance drew
explicit praise: P5 reported that the verbatim excerpts were what made the
written insights credible to them, though P5 rated the overall standard of the
set at 3, attributing this in free text not to inaccuracy but to a circular
restatement and a dangling pronoun in the cross-source synthesis.}

\added{\emph{Utility} was high but uneven. All seven would use the capability in
their own workflow, and six estimated a saving, those giving hours putting it
between five and sixteen, though P5 noted they would probably still spend the
same time reading the papers afterwards, so these are self-reported expectations
rather than measured task times. Whether the insights suggested directions a
participant had not considered tracked their familiarity with the corpus: P6,
the novice and the only participant new to their corpus, rated this item highest
(6), while the two who had helped assemble their corpus rated it lowest (2 and
3). P7, whose corpus consisted entirely of their own recent publications,
nonetheless rated this item 5 and asked for cross-referencing across the corpus.
P6 would use the system but rated recommending it to junior researchers at 2,
citing the difficulty of validating generated insights and of noticing what is
missing, a limitation that applies most sharply to the readers least able to
check the output.}

\added{Across all four groups, every participant independently asked for
explicit signalling of evidential strength: citation lineage, confidence scores,
uncertainty indicators, venue and publication year, cross-referencing across the
corpus, and direct linking of each quoted passage to the claim it supports.}\revnote{SR}

\subsection{Case Study 1: Empirical Validation and Refinement of the Gradient
Wedge Hypothesis}

\added{Both case studies were conducted as 60-minute remote sessions in
which a developer operated the system and the researcher directed the analysis.}\revnote{R2b: study protocol} In
the first, E1 used EVis+ to investigate how air and rock surface temperatures
contribute to rock cracking. E1 uploaded the Geomorphology dataset, a time
series of 23 environmental sensor variables over 26,304 records
\cite{eppes2020warmer}, together with its curated corpus of 31 publications, and
specified the analytical intent: ``Determine the relationship between air
temperature and rock surface temperature as it relates to rock cracking.'' The
system synthesized the corpus into four LD-insights presented as cards in the
insight view.

 Upon
reviewing these cards, E1 focused on the Gradient Wedge insight, which describes
how temperature differences between adjacent rock facets, the exposed
faces of a rock, generate directional tensile stresses that influence crack
orientation. They examined the associated variables and
inspected the source evidence excerpts to verify that the underlying claims
originated from primary empirical findings rather than secondary citations.
However, they noted a discrepancy: while the mechanism was classified as
well-established based on its prevalence across sources, E1's inspection of the
source evidence revealed that much of this support was observational and direct
empirical validation of the proposed stress-driven cracking mechanism remained
limited. This closer reading of the evidence motivated a focused investigation.

\begin{figure} \centering \includegraphics[width=\columnwidth,
alt={}]{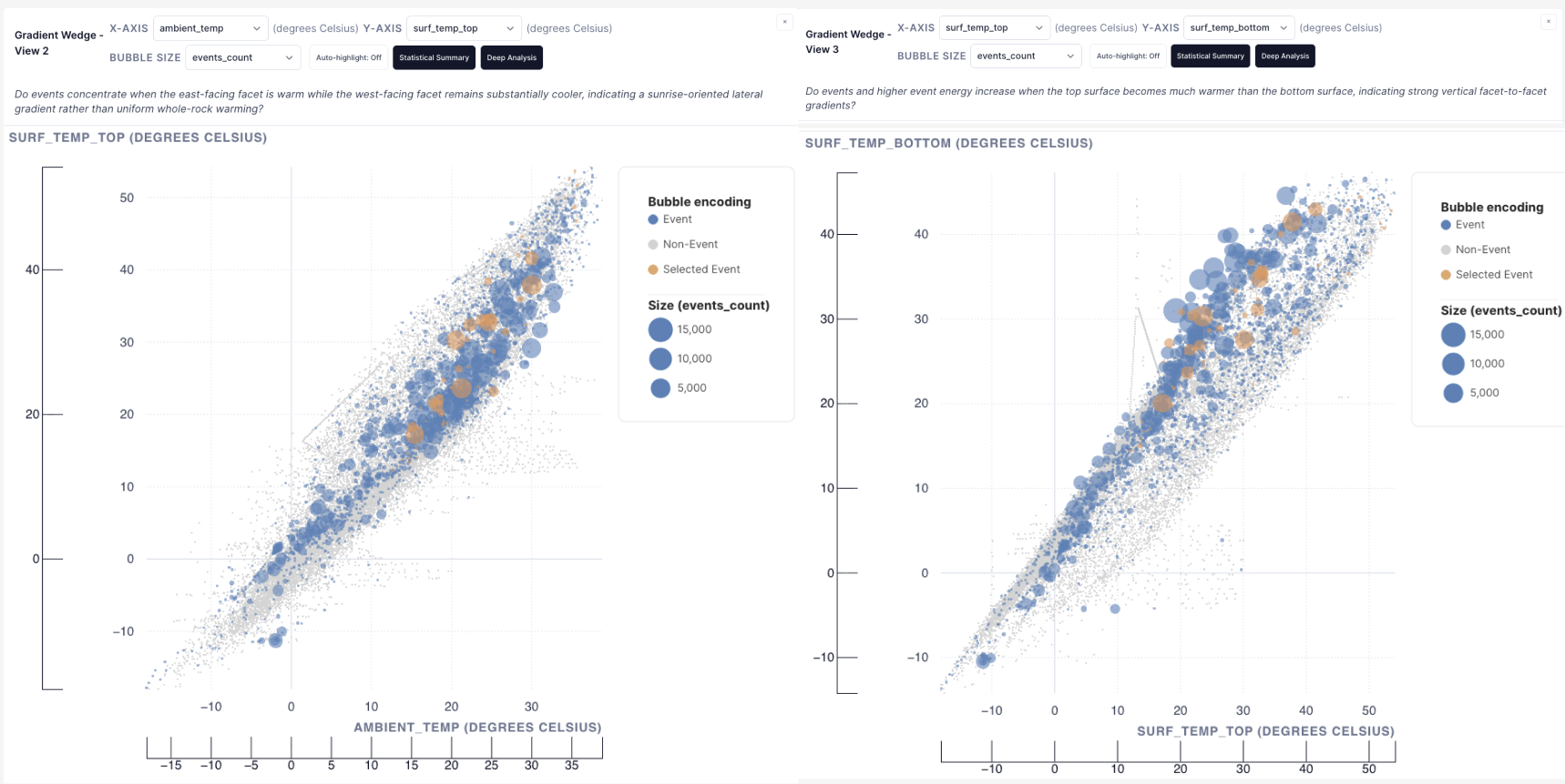} \caption{Case Study 1: Bivariate scatterplots
exploring temperature differentials across rock facets. Left: ambient
temperature vs. top surface temperature, showing higher-energy cracking events
where surface temperature exceeds air temperature. Right: top vs. bottom surface
temperature, revealing elevated cracking activity when bottom surface
temperature exceeds top.}

\label{fig:case_study_1} \vspace{-8pt}

\end{figure}

To begin, they clicked the "Recommend Visualization"
button of that LD-insight and EVis+ showed them an initial
hypothesis: ``If event frequency and energy increase during early-morning
periods when a sun-exposed facet is significantly warmer than its opposing facet
 \ldots{} then the Gradient Wedge mechanism is
active, as directional heating induces facet-to-facet thermal gradients that
generate tensile stress and promote crack initiation.'' E1 inspected the
hypothesis and the associated analysis questions to verify that they aligned
with their understanding of the mechanism. EVis+ also recommended a list of visualizations, including
diurnal time-series plots, bivariate scatterplots, faceted heatmaps, histogram
small multiples, and layered temporal views, and instantiated them after E1
accepted the list. E1 began by examining the diurnal time-series and bivariate
scatterplots to verify the presence of temperature differentials across rock
facets. They tested different temperature variables in
these views by interactively changing the mapping variables. While some evidence
suggested cracking events occur when east--west surface temperatures are
approximately symmetric, as the views suggested, they
observed a stronger pattern when visually exploring the data: cracking events
were more frequent when the bottom surface temperature exceeded the top surface
temperature (Fig. \ref{fig:case_study_1} right).

Further exploration using a diurnal time-series scatterplot revealed a cluster
of events during September 2011. E1 selected this cluster and used the
Statistics View to examine the hourly event rate and energy distributions for
the selected subset, confirming that the cluster represented a genuine
concentration of activity rather than an artifact of data coverage.
 In a bivariate plot of ambient temperature versus top surface
temperature (Fig.~\ref{fig:case_study_1} left),
they observed that higher-energy events tend to occur when surface
temperature exceeds air temperature, in a similar pattern to clusters forming
during other seasons. This pattern suggested a potential decoupling between air
and surface conditions, possibly driven by transient environmental factors such
as sudden wind changes that cool the air while surface temperatures remain
elevated, regardless of absolute temperatures.

\begin{figure}[t] \centering
\includegraphics[width=\columnwidth]{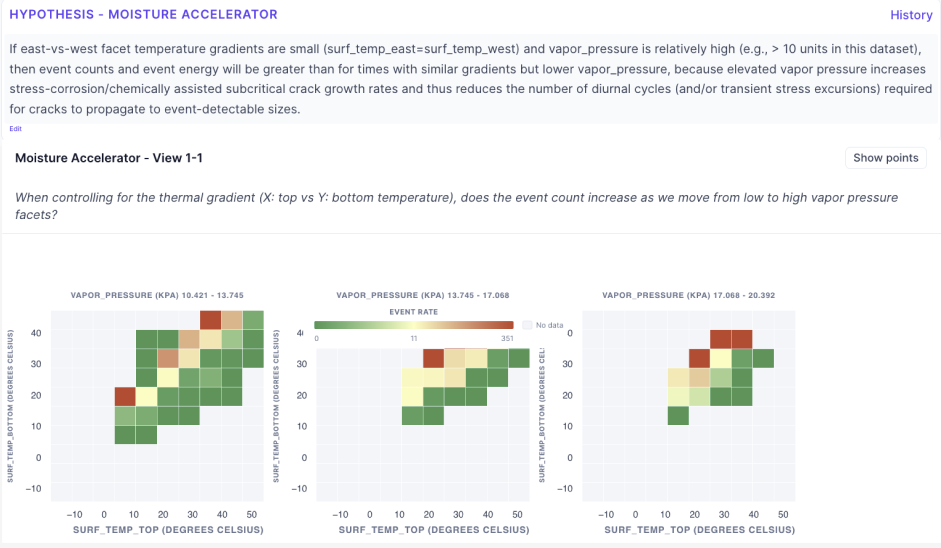} \caption{Case Study
1: the refined hypothesis from the moisture-focused re-exploration (top) and a
faceted heatmap of event rate across top and bottom surface temperatures for
different vapor pressure ranges (bottom), showing increased cracking where vapor
pressure is high and the bottom surface is warmer.}

\label{fig:faceted-heatmap} \vspace{-15pt}

\end{figure}

Based on these observations, E1 refined the hypothesis to focus on
atmospheric moisture effects: ``For periods with similar thermal gradients,
higher vapor pressure is associated with increased cracking activity.'' This
transition from the initial Gradient Wedge hypothesis to a moisture-focused
investigation illustrates the exploratory expansion loop central to the
SciGroundVA workflow. To support this
hypothesis, EVis+ generated a
faceted heatmap (Fig.~\ref{fig:faceted-heatmap}) showing top and
bottom surface temperatures across different vapor pressure ranges. They
revealed that when vapor pressure is high and bottom surface temperature exceeds
top surface temperature, event frequency increases significantly, but declines
rapidly as the temperature differential diminishes. These findings suggest that
 moisture may amplify or modulate
thermally induced stress.

E1 further investigated cases where east--west surface temperatures were
approximately equal, aiming to understand cracking behavior under minimal
directional gradients. Using the deep analysis function,
they triggered EVis+ to retrieve and synthesize additional
 LD-insights relevant to these
 conditions, re-entering the LD-insight
synthesis phase with new analytical context. The system
generated a refined, LD-insight-based hypothesis for another round of visual
exploration.

Through this iterative process, E1 moved beyond the initial Gradient
Wedge hypothesis to an understanding incorporating both thermal gradients and
moisture effects, concluding that cracking events are most prevalent when
thermal imbalance persists across facets and environmental conditions facilitate
propagation. E1 reported having long suspected these relationships without
formally testing them, and that the initial insight list had recalled the idea
that led them to select the Gradient Wedge insight.

\deleted{In addition, E1 noted that LD-insights synthesized by the system were strong, accurate and detailed, and would be especially useful to her for students she wished to `jumpstart' on this subject. She is excited to use EVis+ with several other datasets that she has acquired!}\revnote{SR: overstated evaluation claims}

\subsection{Case Study 2: Seasonal Moisture and Cloud Cover Effects on Rock
Cracking}

In the second case study, E1 and E4 collaboratively investigated how
seasonal variability in temperature and precipitation influences cracking
activity in the Geomorphology dataset. E4 is a researcher with 12 years'
experience on soils, geomorphology, and environmental hazards. Their curiosity
about the potential impact of seasonal rainfall patterns on rock cracking came
from their own research on using soils to understand the impacts of climate
change on western landscapes.

 E4 defined the research goal as: “How does seasonal temperature and
precipitation variability affect cracking rates in the desert southwest?” Using
the same geomorphology corpus, EVis+ synthesized and presented three relevant
LD-insights. Among these, E4 was particularly interested in the wide range of
relative humidity thresholds reported across studies. They
chose to focus on the \textit{Chemical Micro-Saw} mechanism, which describes how
atmospheric moisture weakens chemical bonds at crack tips, enabling fractures to
propagate under low stress associated with temperature changes. Based on
this mechanism, EVis+ generated an initial hypothesis linking relative humidity
above the 35--40\% threshold at which salts absorb water and dissolve to faster
crack growth under everyday temperature cycling. To explore this hypothesis,
EVis+ instantiated coordinated views: a time series scatterplot, a faceted
heatmap, histogram small multiples, and a layered temporal bubble matrix
relating cracking activity to the environment (Fig. \ref{fig:teaser}).

 The experts began by
examining temporal relationships between relative humidity and rock cracking
energy in the recommended time series scatterplot. They observed
recurring seasonal peaks in cracking that align with the monsoon season
(June–September) (Fig. \ref{fig:teaser} top right), a period characterized by
increased rainfall and thunderstorms in the southwestern United States. This
pattern suggested a strong seasonal signal, thus linking a known climate
phenomenon to cracking behavior.

Building on this observation, E4 moved to the
 other recommended views
and examined the joint effects of relative humidity and vapor pressure using the
faceted heatmap (Fig. \ref{fig:teaser} bottom right). The
experts identified distinct differences in cracking event distributions across
several variables, including temperature ranges and temperature changes, when
both humidity and vapor pressure were high. Turning
on the heatmap's scatterplot overlay, E4 observed
that regions with elevated vapor pressure and humidity corresponded to clusters
of higher-energy cracking events.

To validate these findings, E1 proposed a closer examination of vapor
pressure trends, changing the y axis of the time series scatterplot from
relative humidity to vapor pressure. They observed that during the monsoon season, vapor pressure
consistently increases and coincides with clusters of intensified cracking
activity. In addition, E4 noted that there exist some small clusters of events
during November, and inferred a relationship between typical seasonal snow onset
and temperatures dropping, which increases vapor pressure. This reinforced the
hypothesis that atmospheric moisture plays a critical role in accelerating crack
propagation, and suggested to E4
 that the climate changes they have
documented may alter cracking
behavior and in turn affect hazards such as rockfall.

 Motivated by these findings, the
experts refined their investigation and introduced a new hypothesis: “Are
cracking events associated with sudden cloudy conditions, and how do changes in
cloud cover influence cracking behavior?” In response,
EVis+ generated a new visualization tailored to this hypothesis, using
total insolation as a proxy for cloud cover. The resulting time series scatterplot revealed clusters of cracking
events occurring during periods of reduced solar radiation. This pattern is
consistent with cracking at the onset of
storms associated with climate phenomena in summer and winter.

 Both experts emphasized that uncovering these nuanced
relationships would have been extremely time-consuming using traditional
workflows, which require extensive manual comparison of data and literature,
and that EVis+ made collaboration easier by supporting exploration of
data and ideas together in real time.

\subsection{Comparative Expert Evaluation}

\added{We compared EVis+ against a baseline system offering the same view types, displayed side by side and coordinated, without LD-insights or view recommendation: researchers chose and configured every view themselves, for example by changing a scatterplot's y-axis variable.}\revnote{SR: no baseline comparison with EVis}

\added{Two participants took part in the comparative study, each with more than five years of experience in their research area. E5 examined their own microalgae bioreactor dataset (28 variables, 27,536 records) using C-Mic, a corpus prepared by their student. E6, a co-author who had no role in developing either system, examined their own hydrogeology time series (9 variables, 6,672 records) using C-HydGeo, a corpus they assembled themselves that includes four of their own papers.}

\added{Each participant completed an individual 90-minute remote session in which a developer operated the system and the researcher directed the analysis. Sessions included 12 minutes introducing the study and both systems, 48 minutes with the baseline, 15 minutes with EVis+, and 15 minutes of oral feedback. The baseline was used first so participants could learn the shared views and configurations before interpreting LD-insights and recommendations. Because the longer baseline segment included both training and exploration, the comparison reflects how literature grounding changes the route to a finding rather than time-to-discovery.}

\added{With the baseline, E5 observed that increasing light intensity coincided
with decreasing CO$_2$ efficiency and increasing CO$_2$ delta, with a departure
from this pattern at the highest intensities; judging the significance of that
departure depended on E5's own domain knowledge. E5 also noted that such
multivariate views can confuse users without visualization expertise. With
EVis+, E5 encountered literature-derived relationships that had not previously
occurred to them, and reported that the corpus increased the perceived
trustworthiness of the results. E5 expressed strong interest in using EVis+,
while commenting that curating the corpus is time-consuming and suggesting that
it be made more automatic.}

\added{With the baseline, E6 identified a temperature threshold near zero
degrees Celsius associated with the onset of flow, found evidence of a lag time of
several hours between air temperature and flow, and noticed an anomaly in late June when flow
persisted despite falling temperatures, after trying many different
configurations. In EVis+, they selected an LD-insight of interest and met the
same pattern directly in the initially recommended views. E6 described their
usual process as data-first, exploring visualizations to find patterns and then
consulting the literature for context, and hoped EVis+ could also support a
data-first entry point.}\revnote{SR: baseline comparison}

\section{Discussion and Limitations}

\label{sec:discussion} Unlike traditional visual analytics systems that
 emphasize data-first
exploration, the SciGroundVA workflow explicitly incorporates prior scientific
knowledge into hypothesis-driven data analysis. By coupling
 LD-insights with expert-curated
visualization templates, the workflow grounds both hypothesis formulation and
visualization design in domain expertise, reducing reliance on unconstrained LLM
generation. Throughout, LLM agents operate within a constrained pipeline
while human judgment gates every transition and controls interactive
exploration. More broadly,
this work suggests a direction for visual analytics that structures domain
knowledge as a first-class component of the analytical process. Several
limitations remain.

\noindent\textbf{Literature retrieval and synthesis quality.} EVis+
operates over a researcher-curated literature corpus. This provides a bounded
and inspectable evidence space, but the quality and coverage of the resulting
LD-insights depend on corpus composition. Relevant publications absent from the
corpus cannot contribute to the synthesis, while low-quality or outdated sources
may still influence generated insights, and corpus construction imposes
substantial effort on researchers, a concern echoed in our expert feedback.

\added{Our audit traced 92.7\% of generated quoted passages to the corpus.
Although this can be improved by adding an automatic grounding check to the
system, neither provenance nor faithful extraction guarantees that all
assumptions, experimental constraints, or conflicting findings have been
recovered. A grounded passage can still be overstated and generalized beyond the
level the cited paper had actually tested, a failure the grounding audit cannot
detect because the quoted passage is itself present in the corpus. LD-insights
should therefore be treated as candidate interpretations whose applicability
requires researcher verification. Every participant in our assessment asked for
explicit signalling of evidential strength, which the prototype does not yet
provide and which future work should add.}\revnote{R3.2, R4e}

\added{\noindent\textbf{LLM reliability and validation boundaries.} The
constrained agent architecture reduces, but does not eliminate, failures from
open-ended LLM generation. Schema constraints, repair procedures, and
pre-rendering checks described in Section~\ref{sec:prototype} ensure structural
and data-schema compatibility rather than scientific validity: a valid
hypothesis may still oversimplify a mechanism, and an executable visualization
may still be inappropriate for the question. The alignment audit that triggers
literature re-exploration is likewise an LLM heuristic rather than a statistical
test, and selected data items are passed directly to the agent; statistical
summaries of larger selections could provide a firmer basis, and this
re-exploration path requires further evaluation. EVis+ therefore exposes
hypotheses, supporting evidence, and recommended views for inspection before
rendering or saving findings. The prototype also depends on a specific model
provider; preliminary tests with an alternative model produced similar results,
but cross-model evaluation remains open.}\revnote{R1, R3.2: what the checks do and do not guarantee}

\noindent\textbf{Visualization template extensibility.} The use of predefined
visualization templates ensures reliable translation from literature-derived
mechanisms to visual representations, but also constrains the analytical design
space. \added{Authoring them is a one-time cost per data type and task rather than per
study, since the library is reused across datasets of the same kind.}\revnote{R1c: template authoring effort}
 Some analytical tasks
may require visual encodings beyond the current template library, which could
limit generalizability across domains. Future work could address this through
hybrid approaches that combine template-based generation with user-driven
customization.

\added{\noindent\textbf{Evaluation scope and generalizability.} The grounding
audit covers eight literature corpora, and the researcher assessment covers
seven participants working across six corpora in three disciplines. The case
studies examine the workflow in depth with two geoscientists, and the
comparative study provides observations from two researchers in different
disciplines. Together these give complementary evidence about technical
reliability and perceived analytical usefulness, but they do not establish
statistically generalizable improvements in task performance or broad domain
independence. Further studies across additional scientific domains, datasets,
visualization requirements, and researcher populations are needed to assess how
well the workflow transfers beyond the contexts examined here. The assessment
also complicates the expectation that the system helps most where a researcher
is least familiar with the literature: the one novice participant would use it
themselves but rated recommending it to junior researchers lowest of any
participant, citing the difficulty of validating generated insights and of
noticing what is missing. The evaluation also surfaced a design question the
current workflow does not address: it assumes a literature-first entry point,
whereas E6 described working data-first, examining visualizations and consulting
the literature afterwards. Supporting both entry points would let researchers
who already have a clear expectation begin from the data.}\revnote{R2c, SR: generality and scope}

\section{Conclusion}

\label{sec:conclusion} Exploratory analysis of scientific observational data is
often guided by prior scientific knowledge, yet visual analytics systems
rarely integrate that knowledge directly into the analysis process. We
introduced SciGroundVA, a workflow that incorporates empirical findings reported
in the literature into hypothesis-driven visual analysis. By combining
LD-insights with expert-curated visualization templates and agentic AI,
SciGroundVA makes the transition from published evidence to hypotheses and
visualizations explicit and inspectable, while leaving the researcher in control
of which insights and hypotheses to pursue.

We built EVis+, a working prototype that applies this workflow to
condition–event time series analysis in geoscience, and evaluated it through a
grounding audit, a structured assessment of the generated insights, case studies
with geoscientists, and a comparative study. Together these show how
literature-derived mechanisms can direct exploration, and where the generated
insights still require researcher scrutiny. \replaced{We evaluated the generated
LD-insights across three disciplines, but tested the full workflow mainly in
geoscience. We intend to extend EVis+ to further domain applications and to
evaluate the workflow with broader participation.}{Although our evaluation is grounded in geoscience, the underlying architecture is domain-independent, and future work will evaluate SciGroundVA across additional scientific domains and with broader user populations to assess this generalizability.}\revnote{R2c, SR: generality}

\bibliographystyle{abbrv-doi-hyperref}

\bibliography{main}

\clearpage

\appendix % You can use the `hideappendix` class option to skip everything after \appendix
% \crefalias{section}{appendix} % this is to make sure that cleverref switches to referring to Appx. X from here on

\section{Implementation Details, Agent Prompts, and Structured Outputs}
\label{app:implementation}

This appendix supplements Section~5 with the model-facing prompts and principal
structured outputs used by EVis+. Consistent with the current system, EVis+ uses
two role-specific agents: the Scholar Agent for LD-insight synthesis and the
Analysis Agent for hypothesis formulation and visualization recommendation.
Phase 3 remains researcher-driven. This appendix also documents the runtime
configuration, the coordinator-level decomposition, and the deterministic
visualization checks.

\subsection{Runtime Workflow and Tool Configuration}

\addedhere{R4b: agent workflow, call sites, and tool configuration}

Table~\ref{tab:agent-runtime} summarizes every LLM call site in the current
backend. Only the Scholar extraction/repair operations and discrepancy-driven
mechanism discovery are given the \texttt{file\_search} tool. The Analysis Agent
operations reason over structured artifacts passed by the coordinator and do not
independently query the literature.

\begin{table*}[t] \centering \caption{LLM operations in \texttt{backend-v4}.
``None'' means that no external tool is provided to that model call.}
\label{tab:agent-runtime} \small
\begin{tabular}{p{0.21\linewidth}p{0.12\linewidth}p{0.18\linewidth}p{0.20\linewidth}p{0.22\linewidth}}
\hline \textbf{Operation} & \textbf{Model} & \textbf{Tool} & \textbf{Structured
output} & \textbf{Invocation context} \\ \hline Intent clarity check &
GPT-5-nano & None & \texttt{IntentClarity} & Before initial literature retrieval
\\ Scholar Agent: extraction & GPT-5-mini & \texttt{file\_search} &
\texttt{ScholarInsightList} & Research intent + dataset feature names + corpus
vector store \\ Scholar Agent: schema repair & GPT-5-mini &
\texttt{file\_search} & \texttt{ScholarInsightList} & One targeted pass for
insights missing an analytical task or associated variables \\ Analysis Agent:
hypothesis & GPT-5-mini & None & \texttt{Hypothesis} & Structured LD-insight \\
Analysis Agent: visualization recommendation & GPT-5-mini & None &
\texttt{RecommendationOutput} & Hypothesis + LD-insight + dataset summary +
event fields + existing dashboard state \\ Analysis Agent: alignment audit &
GPT-5-mini & None & \texttt{AuditOutput} & Current insights + hypothesis +
selected observations + dataset summary \\ Scholar Agent: mechanism discovery &
GPT-5-mini & \texttt{file\_search} & \texttt{DiscoveryOutput} & Invoked only
after a \textit{Contradictory} audit \\ Analysis Agent: drill-down & GPT-5-mini
& None & \texttt{RecommendationOutput} & Active hypothesis + selected
observations + dataset summary + current dashboard \\ Analysis Agent:
visualization repair & GPT-5-mini & None & \texttt{RepairOutput} & Only for
recommendations flagged by deterministic validation \\ \hline \end{tabular}
\end{table*}

\paragraph{Corpus construction and literature tool.} A literature corpus is
constructed from researcher-provided files with extensions \texttt{.pdf},
\texttt{.txt}, \texttt{.md}, or \texttt{.docx}. The corpus service uploads the
files, creates an OpenAI vector store, attaches the files through a batch
operation, and stores the resulting vector-store identifier in
\texttt{data/<dataset>/vector\_store.json} for reuse. The Scholar extraction,
Scholar repair, and discrepancy-driven mechanism-discovery calls use this vector
store through the \texttt{file\_search} tool. The current backend does not
manually specify retrieval top-$k$, chunk size, or ranking parameters; these
therefore follow the service defaults. For retrieval-enabled calls, file-search
results are also requested from the API so that retrieval behavior can be
inspected and logged.

\paragraph{Dataset context.} Before agent reasoning, the dataset is summarized
into a compact \texttt{data\_summary} object. It contains a list of column names
and, for each column, its inferred data type and (when applicable) standard
deviation, minimum, maximum, up to three sampled values, number of unique
values, and number of missing values. Initial literature synthesis receives the
researcher-defined intent and the available feature names. Visualization
recommendation receives the richer dataset summary rather than the complete
table. During discrepancy-driven re-exploration, brushed observations are added
as \texttt{selected\_points} together with the original dataset summary, current
hypothesis, current insights, and dashboard state; the selected subset augments
rather than replaces the existing analytical context.

\subsection{Coordinator-Level Decomposition}

\addedhere{R4b: how each prompt step is executed}

The initial workflow is decomposed as follows: \begin{enumerate} \item Check
whether the research intent is sufficiently specific. \item Retrieve and
synthesize LD-insights with the Scholar Agent. \item If
\texttt{suggested\_analytical\_task} is empty or
\texttt{associated\_variables\_to\_explore} is empty, run one targeted Scholar
repair call on only those incomplete insights. Re-check the same two conditions
after repair; remaining incomplete insight IDs are recorded, but the current
implementation does not automatically retry or drop them. \item Generate and
attach a structured hypothesis to each retrieved LD-insight. \item When an
insight is explored, generate visualization recommendations from its hypothesis
and the dataset summary. \item Validate the resulting visualization
specifications deterministically; if violations are found, run one targeted
repair call and then apply the same validator again. \end{enumerate}

Discrepancy-driven re-exploration is decomposed into three separate LLM
operations, rather than one monolithic ``deep analysis'' prompt:
\begin{enumerate} \item \textbf{Alignment audit:} classify the selected
observations as \textit{Aligned} or \textit{Contradictory} with the current
hypothesis. \item \textbf{Conditional mechanism discovery:} only for a
\textit{Contradictory} result, search the same literature corpus for an
alternative mechanism and return new ScholarInsight objects. A hypothesis is
then generated for each new insight. \item \textbf{Drill-down recommendation:}
generate one or more complementary visualizations for the active hypothesis and
selected observations, followed by the same deterministic validation/repair
procedure used for the initial dashboard. \end{enumerate}

\paragraph{Hypothesis versus analysis questions.} The current implementation
keeps these representations separate. The Analysis Agent returns an
\texttt{if--then--because} hypothesis together with an atomic analytical intent,
interaction intents, and an analytical justification. It does \emph{not}
generate the dashboard's analysis questions. In its second stage the Analysis
Agent receives the hypothesis and generates a \texttt{primary\_visual\_question}
plus one \texttt{analysis\_question} for each recommended chart. The coordinator
currently precomputes and attaches a candidate hypothesis to each retrieved
LD-insight, while researcher selection determines which insight/hypothesis is
promoted into an active dashboard.

\subsection{Prompt Templates} \label{app:prompts}

The listings are grouped by agent role. The intent-specificity check is a
lightweight pre-retrieval call rather than a separate agent. The Analysis Agent
uses sequential hypothesis-formulation and visualization-recommendation stages;
discrepancy-driven re-exploration reuses the same Scholar and Analysis roles.
Placeholders in braces are populated from the current session state.

\subsubsection{Pre-Retrieval Intent Specificity Check}
\begin{lstlisting}[style=promptstyle] Decide whether the following research
intent is specific enough to drive a scientific literature search. A good intent
names at least one phenomenon, variable, or relationship to investigate. Reject
only intents that are empty, a single generic word ("explore", "analyze", "look
at the data"), or otherwise give the literature search nothing concrete to look
for.

Research intent: "{intent}" Available dataset variables: {data_features}
\end{lstlisting}

\subsubsection{Scholar Agent: LD-Insight Extraction}
\begin{lstlisting}[style=promptstyle] You are an expert Research Analyst. Your
task is to perform deep, structured extraction and cross-source synthesis from
provided materials using a rigorous, stepwise reasoning process.

User Research Intent: {intent} Key Variables / Factors of Interest:
{data_features}

You MUST extract at least 12-20 distinct insights if they exist. Do not
artificially limit the number of insights. Limit each text field to 1-3
sentences unless absolutely necessary. Each insight should include MULTIPLE
variables when applicable. Do not stop at a single dominant variable if others
are present.

Task Instructions Process the provided documents through the following logical
stages: 1. Evidence Filtering. Identify original empirical findings,
observations, or reported outcomes. Group findings across sources that describe
the same underlying pattern, phenomenon, or effect. Strictly exclude "Related
Work" and speculative discussion unless tied to specific results. 2. Causal
Linkage. For each clustered pattern: Identify the underlying drivers,
mechanisms, or explanatory factors (the "why"). Construct a structured reasoning
chain: Observation -> Interpretation -> Underlying Driver / Model. Assign a
plain_language_alias (intuitive or metaphorical label for the concept). 3.
Analytical Task Classification: Categorize the research method (e.g.,
Comparative Analysis, Longitudinal Study, Sensitivity Analysis, Case Study). 4.
Multi-Variable Attribution: Match the identified driver to the Variables of
Interest. Determine numerical thresholds, qualitative triggers, and Relationship
Types. 5. Insight Characterization (Consensus vs. Novelty): Consensus Level:
High (corroborated by multiple independent sources), Medium (supported by two
sources), or Low (novel/outlier observation). Scientific/Disciplinary Standing:
Label as "Established" (standard practice/fact), "Emerging" (gaining traction),
or "Anomalous" (contradictory/edge case).

Return structured insights conforming to ScholarInsightList
(Section~\ref{app:schemas}). \end{lstlisting}

\subsubsection{Scholar Agent: Targeted Structural Repair}
\label{app:scholar-repair}

A Scholar insight is considered structurally incomplete when either
\texttt{suggested\_analytical\_task} or
\texttt{associated\_variables\_to\_explore} is empty. The system performs one
targeted repair pass on only those entries, preserving the remaining content and
insight IDs. Residual incomplete entries are retained and logged. This check
concerns structural completeness, not scientific correctness or source fidelity.

\begin{lstlisting}[style=promptstyle] You previously extracted these scientific
insights, but some are missing required fields. Fill in ONLY the missing fields
for the insights listed below, keeping every other field exactly as given. Use
the same source materials available to you for grounding.

User Research Intent: {intent} Incomplete insights (fix
`suggested_analytical_task` and/or `associated_variables_to_explore` if empty,
leave every other field untouched unless it is also empty):
{incomplete_insights}

Return the corrected insights, one entry per input insight, with the same `id`
values, in the same Output Format used originally (a JSON object with key
"insights"). \end{lstlisting}

\subsubsection{Analysis Agent: Hypothesis Formulation}

The first Analysis Agent stage converts the selected LD-insight into a
structured, testable hypothesis. The second stage, below, maps that hypothesis
to analysis questions and views.

\begin{lstlisting}[style=promptstyle] Transform raw empirical observations from
scientific literature into rigorously structured, testable scientific hypotheses
using atomic analytical intents and interaction intents.

Data observations from literature: {insight}

Task instructions 1. Synthesize a scientific hypothesis: Formulate a testable,
falsifiable hypothesis based on the provided insight. Use a high-precision
"If-Then-Because" structure. 2. Atomic Intent Mapping: Classify the research
objectives into the following categories: - Change_over_time: Investigates
temporal dynamics, longitudinal shifts, progression, acceleration, or decay. -
Distribution: Examines spread, skewness, dispersion, inequality, or frequency
patterns. - Correlation: Tests statistical association between variables
(positive, negative, nonlinear). - Cluster: Identifies natural groupings,
subpopulations, phenotypes, or stratified regimes. - Outlier: Detects extreme
deviations or anomalous cases. - Threshold_detection: Identifies tipping points,
saturation levels, nonlinear inflection boundaries. - Causal_attribution:
Isolates the active causal variable from confounders or background noise. -
Evidence_finding: Searches for confirmatory instances directly supporting a
claim. - Exploration: Probes secondary relationships, unexplained variance, or
emergent structures. 3. Interaction Intents (Analytical Operations): Specify
applicable interaction intents used to interrogate the data: - filter --
restrict dataset by condition, subgroup, or threshold - highlight -- emphasize
significant patterns or anomalies - compare -- contrast groups, conditions, or
time periods - segment -- partition into meaningful subgroups - normalize --
adjust for scale or baseline effects - aggregate -- summarize across units -
drill_down -- move from macro trend to micro-level mechanism - rank -- order by
magnitude or effect size

Return a structured Hypothesis object with fields: hypothesis, atomic_intent,
interaction_intents, analytical_justification. \end{lstlisting}

\subsubsection{Shared Visualization Blueprints} \label{lst:blueprints}
\begin{lstlisting}[style=promptstyle] Visualization blueprints (allowed types)
1. Time Series Scatterplot: - Focus on longitudinal change; correlation between environmental spikes and event clusters. - [X: Datetime | Y: Data Column
(quantitative) | R: Quantitative | Color: Boolean Event] 
2. Diurnal Scatterplot: - Focus: Phase-locking; thermal/solar forcing; midnight-to-midnight periodic patterns. - [X: Date | Y: Hour of day (0:00 - 23:00) | R: Quantitative | Color: Boolean Event] 
3. Faceted Heatmap: - Focus: Multivariate interactions (X, Y, Z); identifying "danger zones" or nonlinear regimes. - Constraint: Must use X, Y, and Z, and their bin_count should be in the range 5-15. - [X: Var A (quantitative) | Y: Var B (quantitative) | Z: Var C (quantitative) | Color: Quantitative (target event)] 
4. Histogram (Small Multiples) A grid of N charts:
- Focus: Comparative range-of-effect analysis; Quantifying frequency bias; determining if events are concentrated in a narrow variable range. - X-axis: A unique Data Column (quantitative) for each sub-plot. - Y-axis: Target event (quantitative) 
5. Standard Scatterplot: - Focus: Direct correlation, bivariate clusters, and physical thresholds. - [X: Quantitative | Y: Quantitative | R: Quantitative | Color: Boolean Event] 
6. Layered Temporal Bubble Matrix:
- Focus: Seasonal structure, intra-month patterns, inter-annual variability, and persistent or anomalous temporal event patterns. - Structure: Two overlaid proportional-symbol layers sharing the same X and Y coordinates. - Constraint: Must specify selected_year when X is Day. - [X: Day or Year | Y: Month | R: Quantitative | Z: Data Column (quantitative) | Color: Boolean Event]
\end{lstlisting}

\subsubsection{Shared Visualization Recommendation Output}
\label{lst:recommendation-format}

For concision, the runtime output block is summarized below; the same fields are
enforced by the structured schema in Section~\ref{app:schemas}.
\begin{lstlisting}[style=promptstyle] RecommendationOutput = {
"primary_visual_question": str, "recommended_visualizations": [{ "id": int,
"plot_type": str, "analysis_question": str, "justification": str | null,
"relevance_score": float, "encodings": {x, y, z?, r?, color?, highlights?} |
null, "small_multiples_suite": [encodings] | null, "expected_pattern": str }],
"complementarity_summary": str } \end{lstlisting}

\subsubsection{Analysis Agent: Visualization Recommendation}

This stage selects from the six expert-curated templates and returns
coordinated views with per-view analysis questions, expected patterns, and
supporting/contrasting conditions.

\begin{lstlisting}[style=promptstyle] ROLE: You are an expert scientific data
visualization analyst. Your goal is to provide empirical proof for a scientific
hypothesis by mapping dataset variables to specific visual encodings,
specifically focusing on the relationship between environmental conditions and
discrete events.

Event Definitions: Events & Instances: "Events" are occurrences of interest
(e.g., rockfalls) measured at specific "Instances" (time periods like minutes,
days, or hours). Conditions: These are the simultaneous environmental states
monitored alongside events (e.g., weather, soil moisture). Event Metrics: Events
can be raw data (absolute temperature) or derived metrics (instances where Temp
> 40C). Energy: A quantitative magnitude assigned to an event. This could be the
value of the dimension itself (the actual temp during an exceedance) or a
secondary metric (e.g., the volume of a rockfall).

INPUT: Scientific hypothesis (already established, do not re-derive it):
{hypothesis} Scholarly data insight: {data_insight} Data summary: {data_summary}
Target event: {event_dim} Boolean event: {is_event_dim} Existing dashboard
visuals (empty if this is the first pass): {current_charts}

Guidelines: You are encouraged to use the same Plot Type multiple times if
mapping it to different variables provides unique evidence. When recommending a
Histogram, you must not provide a single chart. Instead, you must generate a
"Small Multiples" suite: a series of histograms, each with a different x field,
to compare how events are distributed across various variable drivers. If
"Existing dashboard visuals" is non-empty, your task is NOT to recreate the
dashboard. Recommend additional charts that: 1) fill analytical gaps, 2) provide
new evidence pathways, 3) complement (not duplicate) existing charts, and 4)
strengthen hypothesis validation.

Task Instructions: 1. Core Visual Question: Define the single most critical
relationship the viewer must see to validate the hypothesis. 2. Selection
Strategy: Determine how many instances of each blueprint are needed. If the
hypothesis mentions multiple drivers, provide corresponding plots to isolate
each. 3. Analysis Question: For every recommended chart, define a specific
Analysis Question the viewer should ask while looking at the plot to verify the
hypothesis. 4. Surgical & Comparative Highlighting: For every chart, identify
the Support Evidence ("Smoking Gun" range/threshold) and Alternative Evidence
(baseline/null/counter-argument range). 5. Expected Pattern: Describe the visual
trend.

{BLUEPRINTS; Listing~\ref{lst:blueprints}} {RECOMMENDATION_OUTPUT_FORMAT;
Listing~\ref{lst:recommendation-format}} \end{lstlisting}

\subsubsection{Analysis Agent: Alignment Audit for Selected Observations}

During discrepancy-driven re-exploration, this classification is used as an
exploratory routing signal rather than as a statistical test.

\begin{lstlisting}[style=promptstyle] ROLE: You are a Senior Data Scientist
auditing whether a user's selected data points are consistent with an existing
scientific hypothesis and literature insight.

Inputs: Literature insights: {lit_insights} Current hypothesis: {hypothesis}
Selection context (the "Points of Interest"): {selected_points} Data summary:
{data_summary}

TASK: Compare the selected points against the hypothesis and literature
insights. - If the selected points follow the patterns described in the
hypothesis/insight, return status "Aligned". - If the points violate the
hypothesis (e.g., events occurring outside expected thresholds), return status
"Contradictory".

Identify the literature insight id (if any) the selected points most directly
relate to, and explain your reasoning in conflict_analysis either way.

Return an AuditOutput object with: status, matched_insight_id,
conflict_analysis. \end{lstlisting}

\subsubsection{Scholar Agent: Discrepancy-Driven Mechanism Discovery}
\begin{lstlisting}[style=promptstyle] ROLE: You are a Senior Data Scientist and
Research Analyst. The selected data points contradict the current hypothesis.
Perform a deep-dive extraction from the available literature to find a new
explanation for them.

Inputs: Literature insights so far: {lit_insights} Contradicted hypothesis:
{hypothesis} Selection context (the "Points of Interest"): {selected_points}
Data summary: {data_summary}

STRICT DATA CONSTRAINT: Only use variables explicitly listed in the "Data
Summary." Do not invent field names or use generic placeholders.

TASK INSTRUCTIONS: 1. Evidence Filtering: Isolate empirical observations from
the literature that match the conditions of the selected points. Strictly
exclude "Related Work" and "References." 2. Mechanistic Linkage: Isolate the
"Why" (e.g., Subcritical Crack Growth). Assign a Plain-Language Alias (e.g.,
"The Micro-Saw"). 3. Variable Attribution: Map the selected points' variables to
the mechanism. Define the Value Range and Relationship Type (Primary Driver,
Threshold Trigger, etc.). 4. Task Classification: Categorize the required
analysis (Threshold Detection, Causal Modeling, etc.).

Produce one or more new ScholarInsight-shaped entries plus an overall_synthesis
sentence. \end{lstlisting}

\subsubsection{Analysis Agent: Drill-Down Visualization Recommendation}
\begin{lstlisting}[style=promptstyle] ROLE: You are an Expert Scientific
Visualization Analyst. Recommend 1-3 "drill-down" visualizations that isolate
the user's selected points against the background data and either reinforce or
refine the current hypothesis.

Inputs: Current hypothesis: {hypothesis} Selection context (the "Points of
Interest"): {selected_points} Data summary: {data_summary} Existing dashboard
visuals: {current_charts} Target event: {event_dim} Boolean event:
{is_event_dim}

TASK INSTRUCTIONS: 1. Look at the existing dashboard visuals. What is missing to
explain this specific selection? 2. Recommend charts that specifically isolate
the selected points against the background data, focusing on secondary drivers
not emphasized yet. 3. For each chart, define a specific Analysis Question, and
a `justification` explaining exactly how this chart provides NEW evidence,
linking each encoding to a role in the causal chain (driver, response,
threshold, proxy), and why alternative explanations would fail to produce the
same pattern. 4. Identify Support Evidence and Alternative Evidence highlight
ranges for each chart, focused on the selection.

{BLUEPRINTS; Listing~\ref{lst:blueprints}} {RECOMMENDATION_OUTPUT_FORMAT;
Listing~\ref{lst:recommendation-format}} \end{lstlisting}

\subsubsection{Analysis Agent: Visualization Specification Repair}
\begin{lstlisting}[style=promptstyle] You previously recommended these
visualizations, but some violate hard constraints against the actual dataset.
Fix ONLY the charts listed below to resolve their violations. Keep every chart's
`id` the same. Only reference fields that actually exist in the data summary;
only set bin_count within the allowed range for the relevant plot type. Keep
every other field unchanged unless it must change to fix the violation.

Data summary: {data_summary} Violations: {violations} Charts to fix: {charts}

Return the corrected charts only, in the same shape, under
"recommended_visualizations". \end{lstlisting}

\subsection{Structured Output Schemas} \label{app:schemas}

The principal Pydantic-style objects exchanged between the two agents and the
application are summarized below. Ranking fields such as
\texttt{relevance\_score} organize outputs and are not calibrated measures of
scientific confidence.

\begin{lstlisting}[style=promptstyle] IntentClarity = { is_specific: bool,
reason: str }

Hypothesis = { hypothesis: str, atomic_intent: str, interaction_intents: [str],
analytical_justification: str }

SourceEvidence = { quote: str, source_title: str, contribution: str }

RelevantFeature = { variable_name: str, value_range_summary: str,
detailed_threshold_analysis: str, relationship_type: Primary Driver | Accelerant
| Threshold Trigger | Modulating Factor | Inhibitor, interaction_detail_summary:
str, causal_interaction_narrative: str, source_evidence: [SourceEvidence] }

ConsensusNoveltyIndex = { level: High | Medium | Low, scientific_standing:
Established | Emerging | Anomalous, justification: str }

MechanismInfo = { name: str, plain_language_alias: str, summary: str,
detailed_mechanic_breakdown: str, cross_paper_synthesis: str, source_evidence:
[SourceEvidence], consensus_novelty_index: ConsensusNoveltyIndex }

ScholarInsight = { id: int, mechanism_info: MechanismInfo,
suggested_analytical_task: str, associated_variables_to_explore:
[RelevantFeature], overall_synthesis: str, relevance_score: float }
ScholarInsightList = {insights: [ScholarInsight]}

EncodingChannel = { field: str, type: str, bin_count: int | null, scale_type:
str | null, selected_year: int | null } HighlightChannel = { field: str, value_range: [float, float],
rationale: str } ChartEncodings = { x: EncodingChannel, y: EncodingChannel, z:
EncodingChannel | null, r: EncodingChannel | null, color: EncodingChannel |
null, highlights: {support_evidence: HighlightChannel | null,
alternative_evidence: HighlightChannel | null} | null }

ChartRecommendation = { id: int, plot_type: str, analysis_question: str,
justification: str | null, relevance_score: float, encodings: ChartEncodings |
null, small_multiples_suite: [ChartEncodings] | null, expected_pattern: str }
RecommendationOutput = { primary_visual_question: str,
recommended_visualizations: [ChartRecommendation], complementarity_summary: str
} AuditOutput = { status: Aligned | Contradictory, matched_insight_id: int |
null, conflict_analysis: str | null } DiscoveryOutput = { new_insights:
[ScholarInsight], overall_synthesis: str } \end{lstlisting}

\subsection{Deterministic Visualization Validation and Repair}

\addedhere{R1, R3.2: non-LLM checks applied to generated specifications}
\label{app:validation}

After initial or drill-down visualization generation, the coordinator validates
the recommendations against the current \texttt{data\_summary}. The current
validator performs the following checks: \begin{enumerate} \item Every
referenced \texttt{x}, \texttt{y}, \texttt{z}, \texttt{r}, and \texttt{color}
field must occur in the dataset's column list. \item Support- and
alternative-evidence highlight fields must exist. If a numeric min/max is
available for a highlighted field, the requested highlight interval must overlap
the observed data range. \item A \texttt{histogram\_small\_multiples}
recommendation must contain a non-empty \texttt{small\_multiples\_suite}; each
sub-encoding is checked using the same field and highlight rules. \item Every
non-histogram recommendation must contain an \texttt{encodings} object. \item A
\texttt{faceted\_heatmap} must contain \texttt{x}, \texttt{y}, and \texttt{z}
encodings. In the current deterministic validator, \texttt{z.bin\_count} must be
within $[5,15]$. \end{enumerate}

If one or more violations are found, only the flagged charts are sent to the
visualization-repair prompt together with the violations and current data
summary. The coordinator performs one repair pass and then runs the same
deterministic validator again. The pre- and post-repair violation counts are
recorded. These checks test structural and dataset-schema compatibility; they do
not establish that a recommended analysis is scientifically appropriate.

\paragraph{Prompt-level versus deterministic constraints.} The visualization
blueprint text additionally asks for bin counts in the range 5--15 for all
dimensions of a faceted heatmap and asks the Layered Temporal Bubble Matrix to
provide \texttt{selected\_year} when its X encoding is Day. In the current
\texttt{backend-v4} code, these two requirements are prompt-level constraints;
the deterministic validator explicitly enforces the faceted-heatmap X/Y/Z
presence and the Z-bin constraint, but does not separately enforce every
blueprint-specific semantic condition.

\section{Additional Figures}

\begin{figure*}[t] \centering
\includegraphics[width=\linewidth,height=0.9\textheight,keepaspectratio]{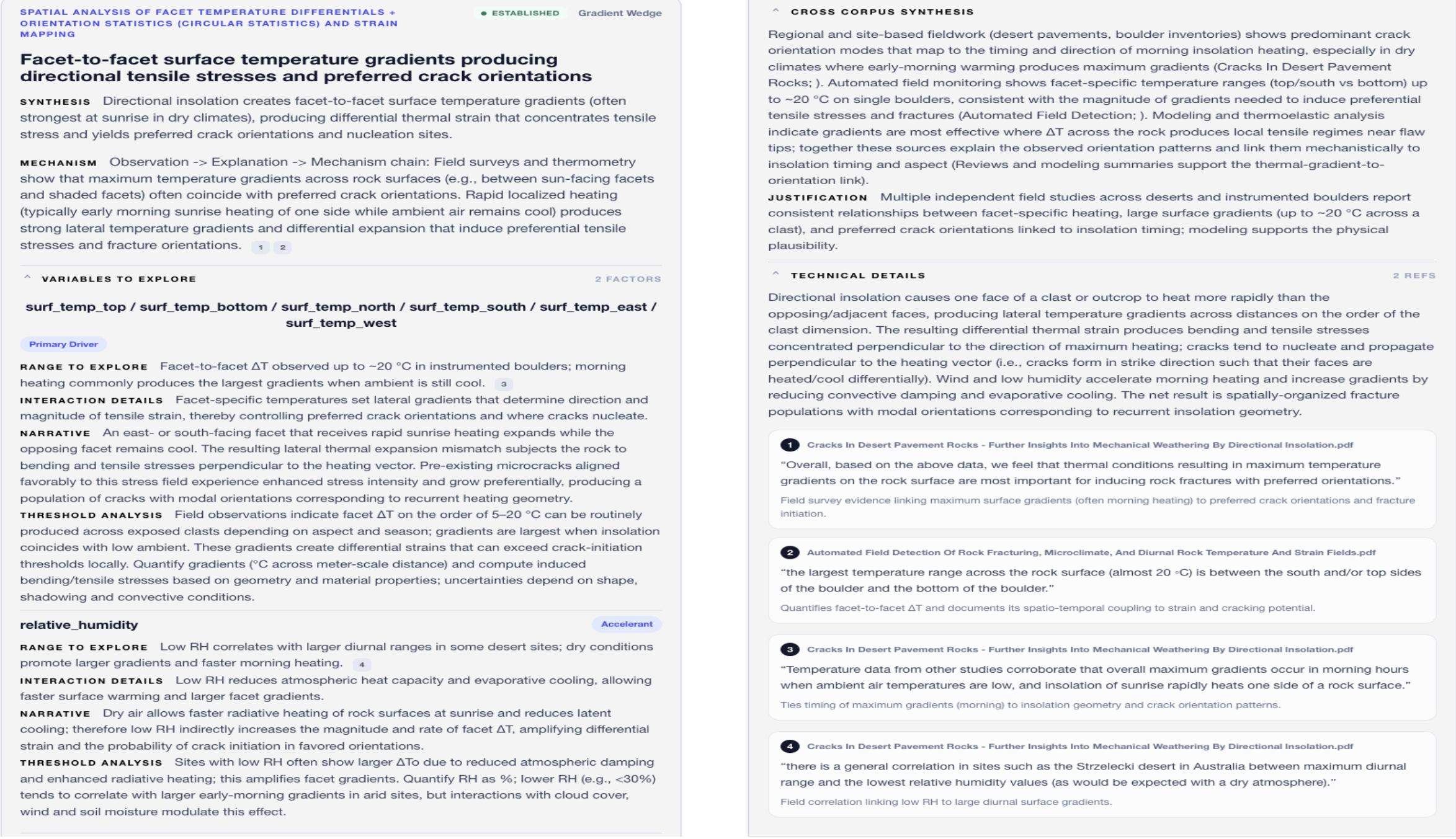}
\caption{The \textit{Gradient Wedge} insight explored in Case Study 1. It
describes how temperature differences between adjacent rock facets generate
directional tensile stresses that influence crack orientation.}
\label{fig:appendix-1} \vspace{-15pt} \end{figure*}

\begin{figure*}[t] \centering
\includegraphics[width=\linewidth]{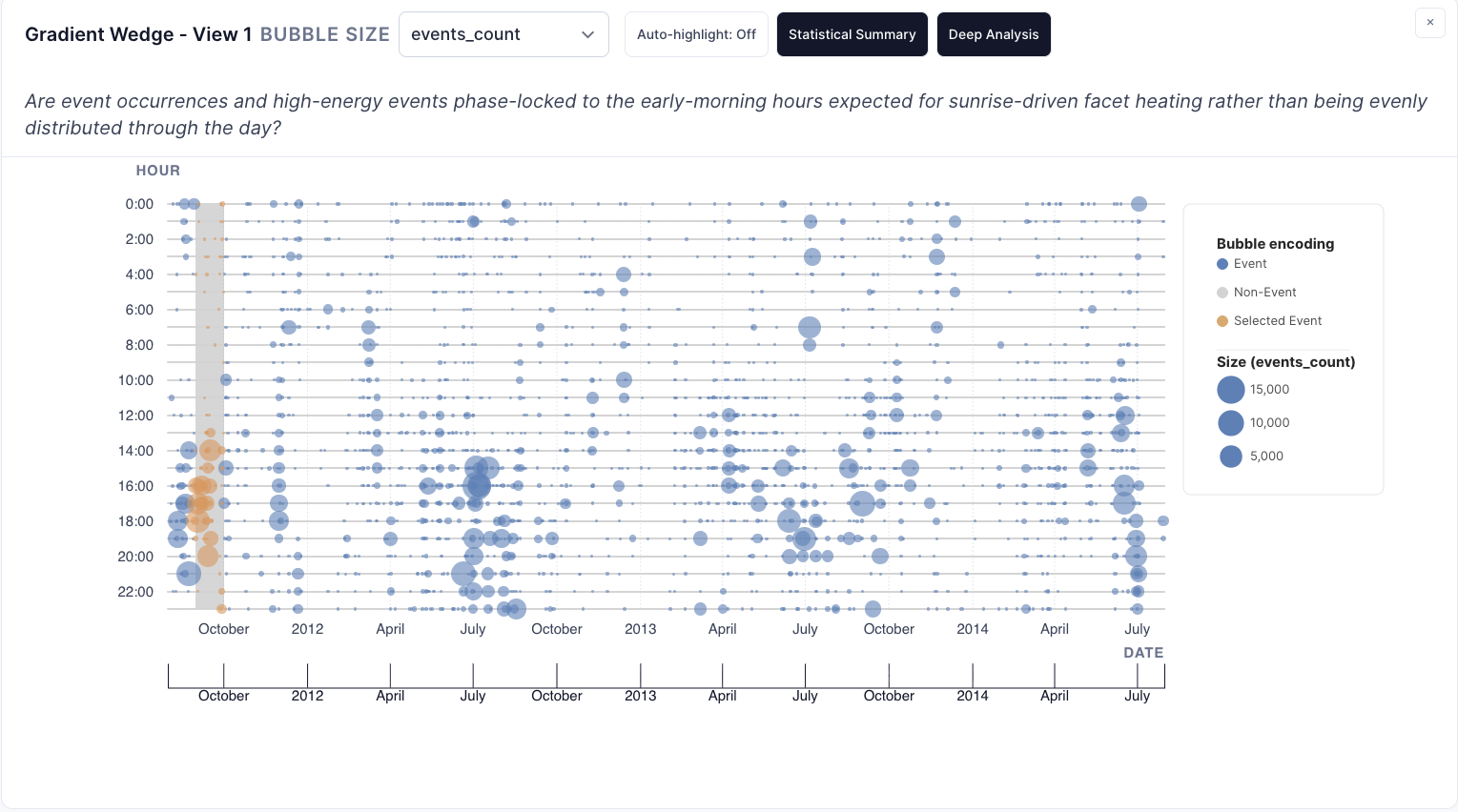} \caption{A diurnal
scatterplot depicting rock cracking event occurrences over multiple years. The
selected points correspond to September 2011 (Case Study 1), during which a
pronounced cluster of elevated cracking activity is observed.}
\label{fig:appendix-2} \end{figure*}

\begin{figure*}[t] \centering
\includegraphics[width=\linewidth]{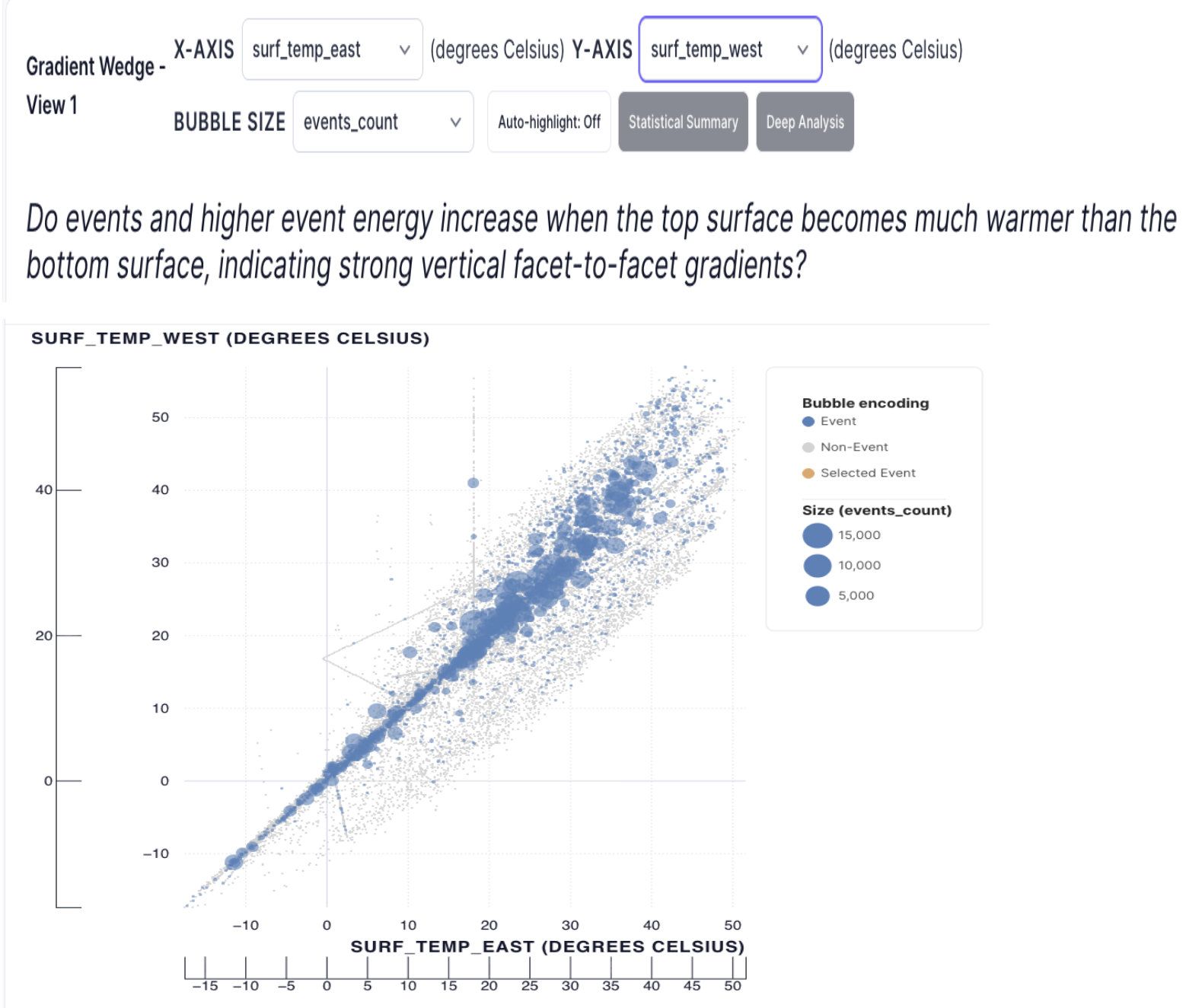} \caption{A scatterplot
of east versus west surface temperature for Case Study 1, where the expert
identifies a cluster of increased rock cracking events occurring when the two
temperatures are nearly equal.} \label{fig:appendix-3} \end{figure*}

\begin{figure*}[t] \centering
\includegraphics[width=\linewidth,height=0.9\textheight,keepaspectratio]{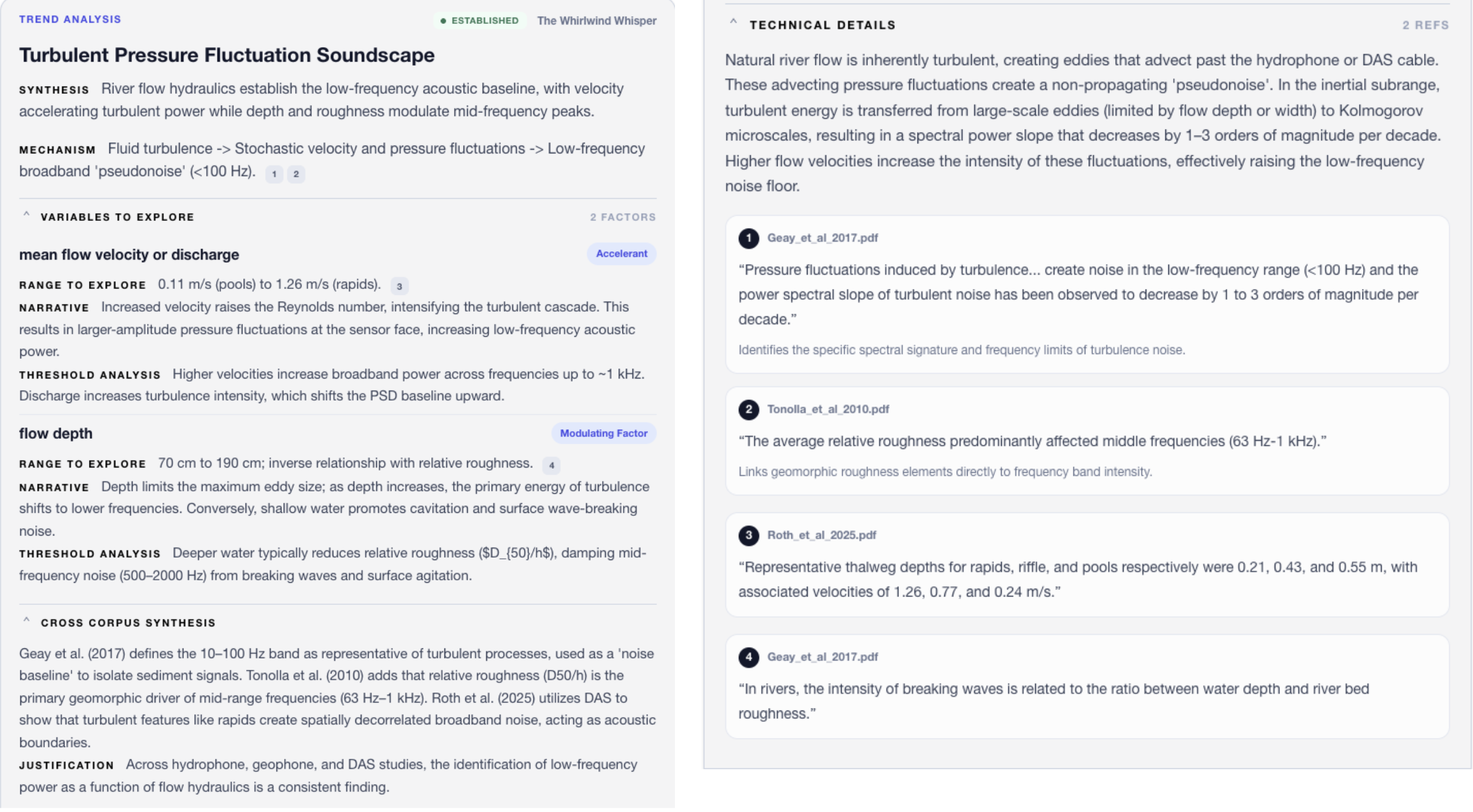}
\caption{An example insight derived from E2's literature collection for an
investigation on relationships between hydroacoustic time series—captured via
hydrophones or distributed acoustic sensing (DAS)—and river flow hydraulics
(flow depth, velocity, turbulence), as well as bedload sediment transport (flux
and characteristic grain sizes). These processes produce acoustic signals with
distinguishable frequency characteristics and power spectral density
signatures.} \label{fig:appendix-4} \end{figure*}

\begin{figure*}[t] \centering
\includegraphics[width=\linewidth,height=0.9\textheight,keepaspectratio]{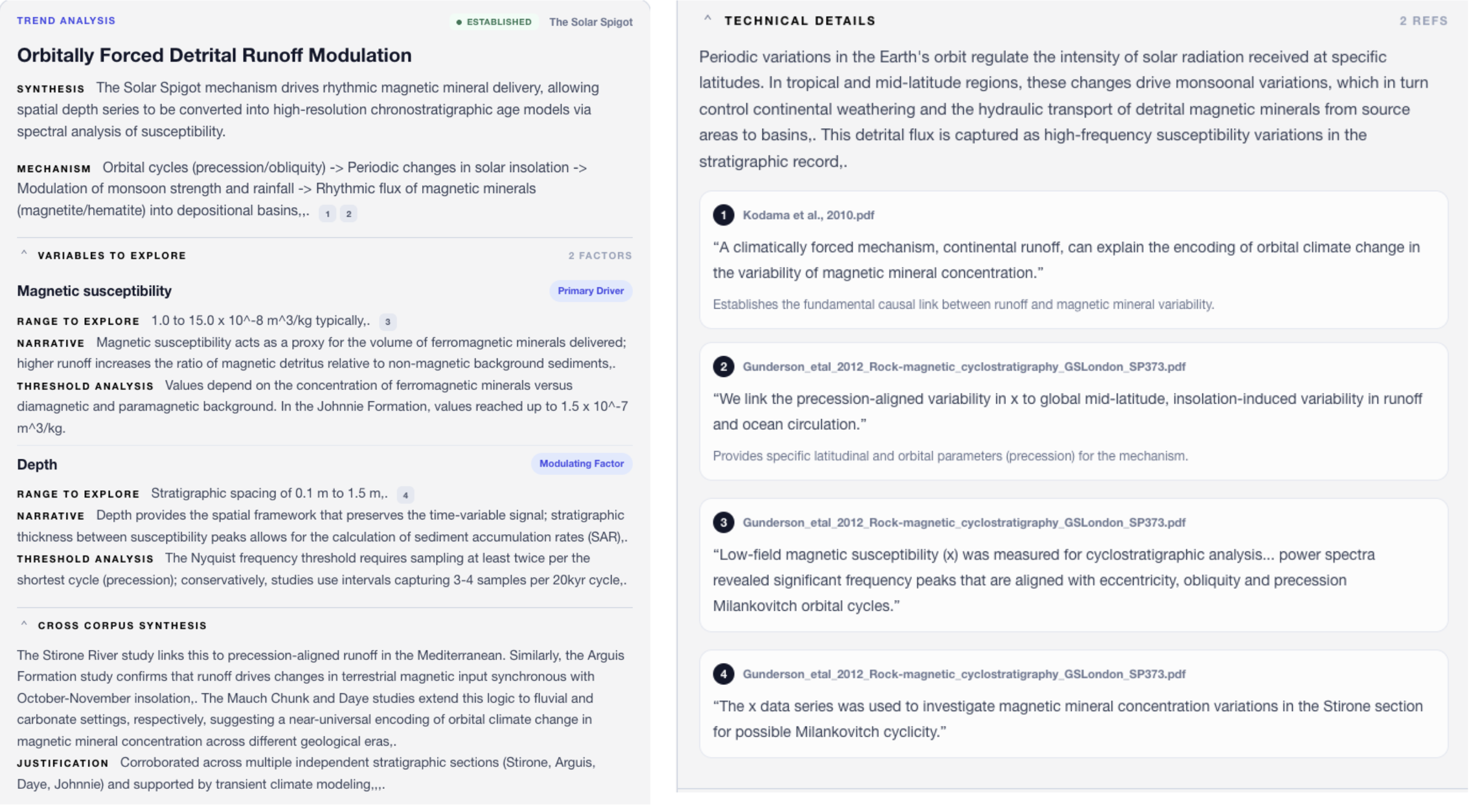}
\caption{An example insight derived from E3's literature collection for the
study of whether variations in rock magnetic susceptibility preserve
Milankovitch-scale cyclicity within sedimentary sequences.}
\label{fig:appendix-5} \end{figure*}

This section presents additional figures that complement the main paper. Figs.
\ref{fig:appendix-1}–\ref{fig:appendix-3} show screenshots of example
LD-insights and visualizations from Case Study 1. Figs.
\ref{fig:appendix-4}–\ref{fig:appendix-5} present screenshots of example
LD-insights derived from E2 and E3’s literature collections, respectively.

\end{document}